\documentclass[journal=jacsat,manuscript=article]{achemso}

\usepackage[version=3]{mhchem} 

\author{Ian Babich}
\altaffiliation{These authors contributed equally}
\email{ian.babich.a@gmail.com}
\author{Andrei Kudriashov}
\altaffiliation{These authors contributed equally}
\author{Denis Baranov}

\author{Vasily Stolyarov}
\affiliation{Advanced Mesoscience and Nanotechnology Centre, Moscow Institute of Physics and Technology, 141700 Dolgoprudny, Russia}
\alsoaffiliation{Superconducting and Quantum Technology Lab, Dukhov Research Institute of Automatics (VNIIA), Moscow, 127055, Russia}

\title[An \textsf{achemso} demo]
  {Limitations of the current-phase relation measurements by an asymmetric dc-SQUID }

\abbreviations{IR,NMR,UV}

\begin{document}


\begin{abstract}
  Exotic quantum transport phenomena established in Josephson junctions (JJs) are reflected by a non-sinusoidal current-phase relation (CPR). The solidified approach to measure the CPR is via an asymmetric dc-SQUID with
a reference JJ that has a high critical current. We probed this method by measuring CPRs of hybrid JJs based on a 3D topological insulator (TI) Bi$_2$Te$_2$Se with a nanobridge acting as a reference JJ. We captured both highly skewed and sinusoidal critical current oscillations within single devices which contradicts the uniqueness of the CPR. This implies that the widely used method provides inaccurate CPR measurement and leads  to misinterpretation. It was shown that the accuracy of the CPR measurement is mediated by the asymmetry in derivatives of the CPRs but not in critical currents as was previously thought. We provided considerations for an accurate CPR measurement that encourage future experiments with reference CPRs different from those that were used before. 
\end{abstract}

\newpage

In the last decades, two main approaches to measure the current-phase relation (CPR) were solidified: methods based on rf-SQUIDs\cite{troeman2008temperature, spanton2017current, hart2019current, frolov2004measurement, sochnikov2013direct, sochnikov2015nonsinusoidal,golubov2004current, haller2022phase}, and on dc-SQUIDs \cite{kayyalha2020highly, li2019zeeman, kayyalha2019anomalous, nichele2020relating, nanda2017current, della2007measurement, lee2015ultimately, murani2017ballistic,ginzburg2018determination}. While rf-SQUID method requires a complicated setup to inductively couple SQUID to the pick-up loop and field coil, the dc-SQUID technique has a rather simple implementation. 

There are two types of dc-SQUID methods distinguished. 
The first one relies on the reference JJ with a known sinusoidal CPR \cite{ginzburg2018determination}.

The second one fully relies on high asymmetry of critical currents of the studied JJ and of the reference JJ $I_c^{JJ}\ll I_c^{REF}$ \cite{kayyalha2020highly, li2019zeeman, kayyalha2019anomalous, nichele2020relating, nanda2017current, della2007measurement, lee2015ultimately, murani2017ballistic}. 
This method will be further referred to as an asymmetric dc-SQUID technique. 

In case a high asymmetry is provided, it is thought that $I_c^{SQUID}(H)$ dependence should directly reflect the CPR of the studied JJ
, thus no mathematical processing is required to extract the underlying CPR. 
The CPR obtained this way will be further called "expected" (ECPR).
In the original paper \cite{della2007measurement}, where this method was developed, Superconductor-Insulator-Superconductor (SIS) was used as a reference JJ. 
Afterwords, this method was adapted for different types of reference JJs such as JJs of the same nature as those under study \cite{ nichele2020relating, nanda2017current} and nanobridges \cite{kayyalha2020highly, li2019zeeman, kayyalha2019anomalous, murani2017ballistic}.

However, we show that the high asymmetry of critical currents is insufficient for accurate CPR measurements in many cases. 
Instead, a high asymmetry in derivatives of supercurrents is criterion for an accurate CPR measurement. 
We validate the claim experimentally and theoretically by studying asymmetric SQUIDs that consist of superconducting Nb nanobridges and Nb-(Bi$_2$Te$_2$Se)-Nb Josephson junctions. 

\begin{figure}[h!]
	\begin{center}
		\includegraphics[width=16cm]{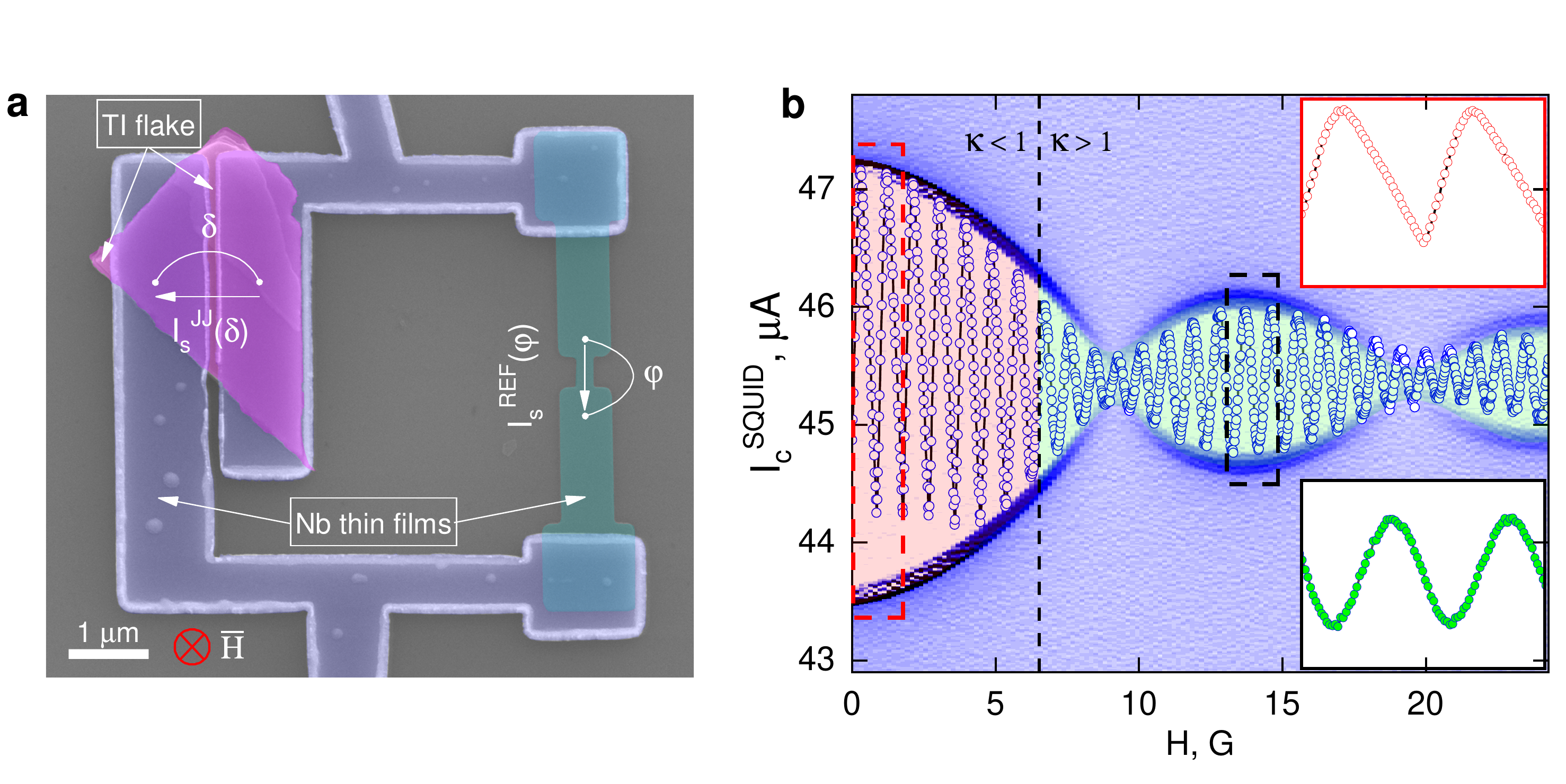}
		\caption{\textbf{ECPR transformation from skewed to sinusoidal oscillations.} (a) A false-colored SEM image of the main device. The TI flake (magenta) forms an S-TI-S JJ carrying $I_s^{JJ}(\delta)$ supercurrent with phase difference $\delta$. The superconducting nanobridge (turquoise) carries $I_s^{REF}(\varphi)$. All the phase differences are defined as the phase on the tip of the arrow minus the phase on the beginning of the arrow. The magnetic field direction is indicated by the red cross. (b)  SQUID critical current oscillations (blue circles) matched with a Fraunhofer pattern (blue-toned gradient) of an S-TI-S JJ with the same material (see Figure S3 of Supprting Information). Red and green regions denote "False" and "True" CPR measurement regimes, respectively (see main text for details). The regions are separated by a black dashed line, which corresponds to $\kappa=I_c^{REF}/I_c^{JJ}\varphi_c=1$. The top and bottom inset graphs are zooms of red and black boxes correspondingly, they display fine features of the oscillations from both regions. 
		}
		\label{Figure1}
	\end{center}
\end{figure}

All of the fabricated devices show universal behavior, thus we will further describe the main device (see Figures S7-S9 of Supporting Information for other samples). An SEM image of the device is presented in Figure \ref{Figure1}a. It is a $5\mu$m$\times 5 \mu $m SQUID loop, where the 100 nm thick Nb film was deposited on top of the TI flake with the distance between the superconducting electrodes of $d=140\;nm$ and a Nb nanobridge with the film thickness of $t=20\;nm$, width $w=230\; nm$ and length $L=390\; nm$.

The critical current of the SQUID as a function of the applied magnetic field is shown in Figure \ref{Figure1}b by blue circles. 
The critical current demonstrates oscillations with the period that is consistent with the magnetic flux quantum $\Phi_0$ per area of the SQUID loop.  
Since a high asymmetry of critical currents is established in the SQUID ($I_c^{REF}\approx 45\;\mu A,\;I_c^{JJ}\approx 1.7\; \mu A$), we should directly observe the CPR $I_s^{JJ}(\delta)$ of the studied junction, i.e., the SQUID oscillations should coincide with the real CPR \cite{della2007measurement}. 

Surprisingly, we observed the transformation of the critical current oscillations shape with the increasing magnetic field. 
We distinguished two completely different behaviors which are labeled by red and green regions in Figure \ref{Figure1}b.
In the red region, the oscillations are skewed, contain linear segments (see top inset graph in Figure \ref{Figure1}b), and do not follow a symmetric envelope of a single JJ Fraunhofer pattern (dark-blue line in Figure \ref{Figure1}b). 
However, the skewness gradually decreases and the oscillations become sinusoidal when the magnetic field exceeds a 7 G value (see bottom inset graph in Figure \ref{Figure1}b). 

A key feature of the measured oscillations in the red region is that there are sharp corners in the minimums, whereas the maximums are smooth (top inset graph in Figure \ref{Figure1}b). 
Similar behavior is observed in several works \cite{kayyalha2020highly, li2019zeeman, della2007measurement, lee2015ultimately, murani2017ballistic}, where this feature was either not discussed or was thought to be caused by exotic properties of the studied object.
Since we would expect preserved time-reversal symmetry in an S-TI-S JJ, the equality $I_s(\delta)=-I_s(-\delta)$ should hold \cite{golubov2004current}. 
Nevertheless, such a non-antisymmetrical \cite{kayyalha2020highly} shape of the measured oscillations does not allow the supercurrents to map onto themselves upon simultaneous change of signs in the current and phase, thus the equality is violated.

Since the oscillations in the red region are significantly different from those in the green region, we raise the question how ECPRs in the red and green regions are related to the real CPR of the studied JJ.

The ambiguity of the experiment interpretation may be lifted after careful revision of the usually briefly discussed theory. The full analysis may be found in Supporting Information. Below we show the core relation following from the analysis:

    \begin{equation}
    \label{eq4}
        \partial_{\varphi}I_s^{REF}(\varphi) +\partial_{\varphi} I_s^{JJ}(\varphi+2\pi\Phi/\Phi_0) = 0,
    \end{equation}
    where $I_s^{REF}$ and $I_s^{JJ}$ are the supercurrents of the reference and the studied JJ correspondingly. The solution of this equation $\varphi^*(\Phi)$ is the phase difference over the reference JJ that maximizes the supercurrent in the SQUID. In case the root is $\varphi^*\approx\varphi_c$ (critical phase of the reference JJ) the SQUID oscillations will be a vertically shifted studied CPR. It is commonly assumed that the asymmetry in critical currents is enough for such situation, however it is already can be seen that the solution is affected by both critical currents and shapes of the used CPRs.

Therefore, we introduce "True" and "False" CPR measurement regimes, where the former corresponds to the case of $\varphi^*(\Phi)$ well localized near $\varphi_c$ providing accurate measurements. The latter corresponds to the SQUID oscillations with CPR mixing, where the ECPR extracted conventionally does not coincide with the underlying CPR.

We solve the equation (\ref{eq4}) for our system, assuming that the studied CPR is sinusoidal $I_s^{JJ}(\delta) = I_c^{JJ}\sin(\delta)$.
It is known, that in the limit of extremely low temperatures and high lengths $L$ of the junctions ($L/\xi\gg1$, where $\xi$ is the coherence length in the weak link), nanobridges show multivalued CPR with linear shape $I_s^{REF}(\varphi) = I_c[(\varphi-\varphi_c \mod{2\pi})+\varphi_c-2\pi]/\varphi_c$, where $\varphi_c$ may be tens of $\pi/2$ \cite{murphy2017asymmetric, vijay2009optimizing, dausy2021impact} (see Supporting Information for side experiments determining nanobridges' CPR).

\begin{figure}[h!]
		\begin{center}
			\includegraphics[width=16cm]{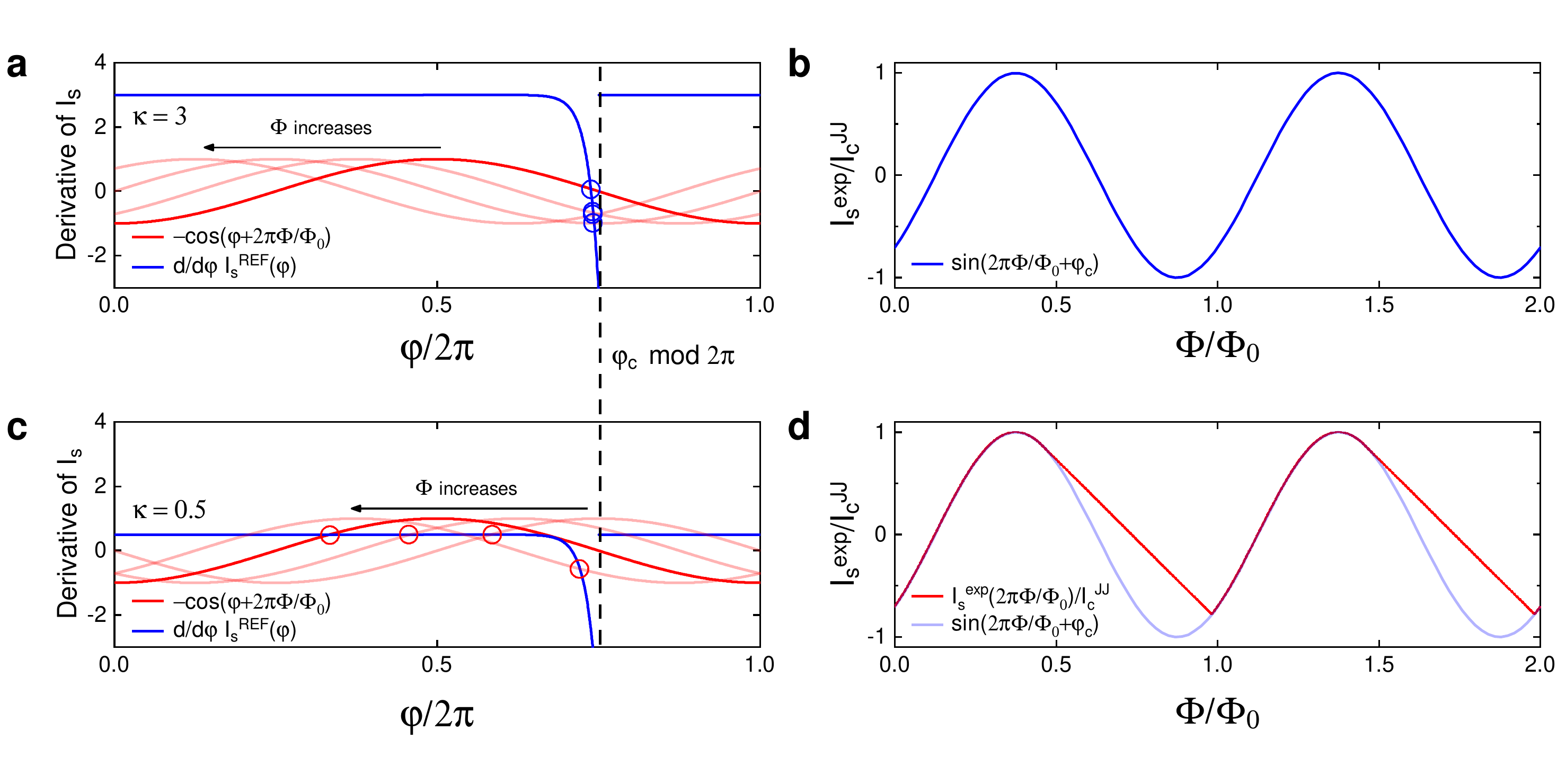} 
		\caption{\textbf{Model for an asymmetric SQUID with a nanobridge.} (a) A graphical representation of the equation (\ref{eq5}) for a "True" CPR regime.  The blue line indicates a realistic derivative of the reference nanobridge junctions CPR for $\kappa =I_c^{REF}/I_c^{JJ}\varphi_c = 3$.
		We depicted a singular behavior near the critical phase to account for the near vertical segments of the sawtooth shaped CPR of the nanobridge.
		Red lines indicate negative derivatives of the studied sinusoidal CPR, that slide to the left with the applied magnetic field. Blue circles denote solutions $\varphi^*(\Phi)$. Black dashed line locates $\varphi_c\mod 2\pi$.  (b) ECPR obtained from an asymmetric SQUID with a nanobridge when $\kappa=3$. (c) A graphical representation of the equation (\ref{eq5}) for the "False" CPR regime.  The blue line indicates a realistic derivative  of the reference nanobridge junctions CPR for $\kappa =0.5$. 
			Red lines are negative derivatives of the studied sinusoidal CPR at different flux values. Red circles indicate solutions $\varphi^*(\Phi)$ that maximize the SQUID supercurrent. (d) The red curve shows the ECPR for $\kappa=0.5$ that does not coincide with the underlying sinusoidal CPR (blue).
			}
			\label{Figure2}
		\end{center}
	\end{figure}

Introducing a dimensionless parameter $\kappa = I_c^{REF}/I_c^{JJ}\varphi_c$, we may write 
 the condition (\ref{eq4}) in the following form:
\begin{equation}
    \kappa(1-\operatorname{\delta}(\varphi-\varphi_c+2\pi n)) = -\cos(\varphi +2\pi\Phi/\Phi_0),
    \label{eq5}
\end{equation}
where $\delta(\varphi-\varphi_c)$ is the Dirac delta-function and $2\pi n $ term arises from nanobridges CPR periodicity. The solution $\varphi^*(\Phi)$ of equation (\ref{eq5}) can be graphically represented as an intersection of two curves, that stand for CPRs derivatives. One of the curves is a negative cosine (right hand side of the equation (\ref{eq5})), and the other is a constant with a singularity (left hand side of the equation (\ref{eq5})). There are two different operating regimes of the system: $\kappa<1$ and  $\kappa>1$ ("False" and "True" CPR regimes correspondingly).
In Figure \ref{Figure2}a, we show the "True" CPR case for $\kappa=3>1$, where the solutions of the equation (\ref{eq5}) (blue circles) are well localized near the critical phase $\varphi_c$ and do not leave its origin with the applied magnetic flux. 
In this case the ECPR coincides with the real CPR (see Figure \ref{Figure2}b). 
The opposite situation is realized for $\kappa<1$, as shown in Figure \ref{Figure2}c. 
In this case, the constant level has more intersections with the negative cosine (red circles), giving rise to new solutions that are not localized near the critical phase. This results in a significant deviation of the ECPR from the underlying CPR, as shown in Figure \ref{Figure2}d. 
We expect the critical phase not to change with the applied magnetic field, thus $\kappa$ is proportional to the critical currents ratio and rises with the decaying amplitude of Fraunhofer oscillations, which explains ECPR transformation in Figure \ref{Figure1}b.

\begin{figure}[h!]
		\begin{center}
			\includegraphics[width=16cm]{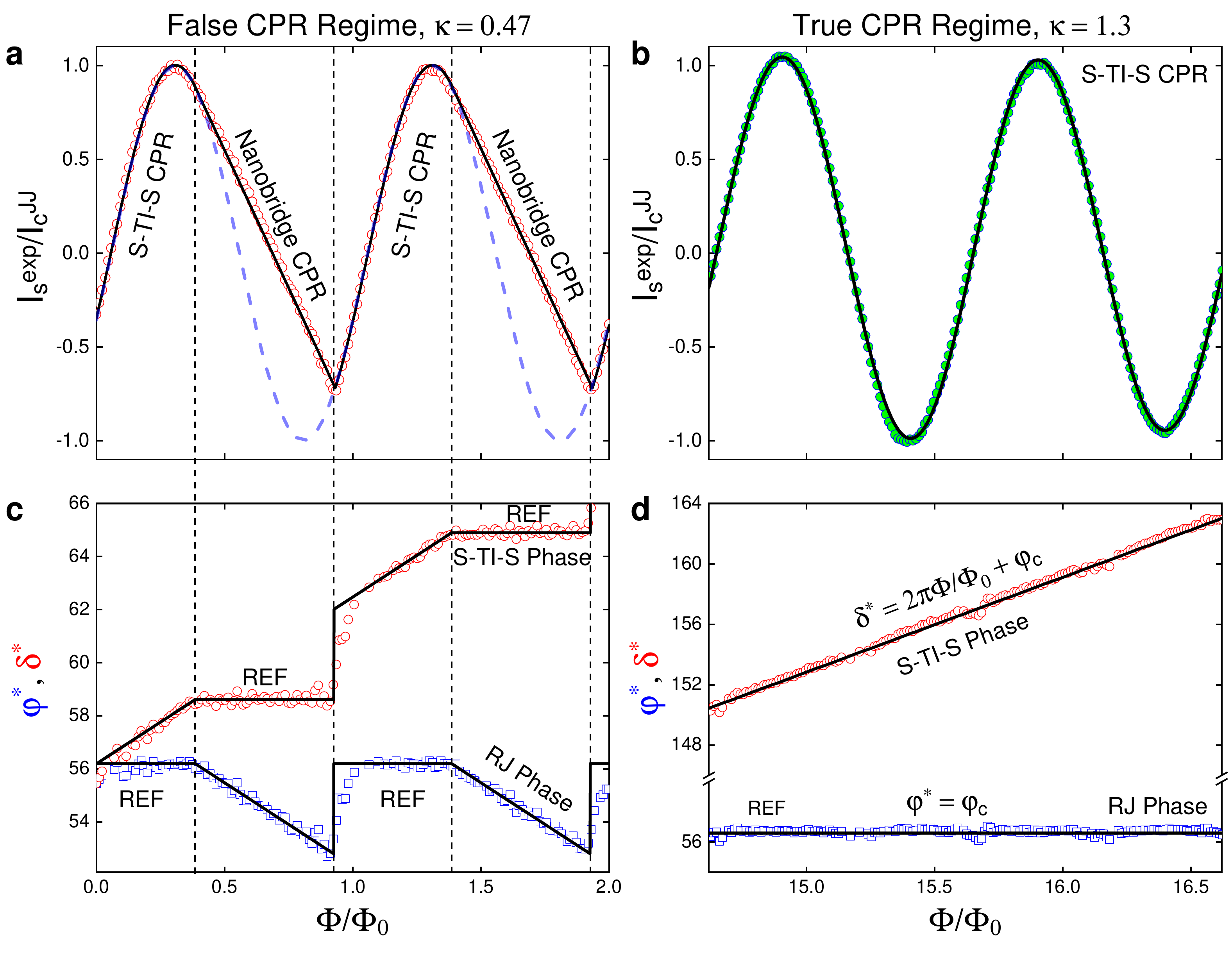} 
			\caption{\textbf{Analysis of the captured CPR measurement regimes.}
   (a) Precise SQUID oscillations measurement near the zeroth maximum (red box in Figure \ref{Figure1}b). The black curve is the fit with the model. The experimental data is contrasted with the sinusoidal curve (blue dashed line). Linear parts of the oscillations attribute to the nanobridges' CPR, whereas sinusoidal parts correspond to the real S-TI-S CPR. (b)  Precise measurement of the SQUID oscillations near the first Fraunhofer maximum (black box in Figure \ref{Figure1}b). The black curve is the fit with a sine. (c) Extracted phases $\varphi^*(\Phi)$ (blue) and $\delta^*(\Phi)$ (red) that maximize the supercurrent of the SQUID. In (a) the black curves display the phases provided by the model model. Regions of constant phases match with the corresponding segments of the CPRs in (a). (d) Phases $\varphi^*$ and $\delta^*$ extracted from (b). The phase of the reference JJ is localized near the critical phase $\varphi_c\approx 56$. The phase of the studied JJ obeys a linear law.
			}
			\label{Figure3}
		\end{center}
	\end{figure}
	
\begin{figure}[h!]
		\begin{center}
			\includegraphics[width=16cm]{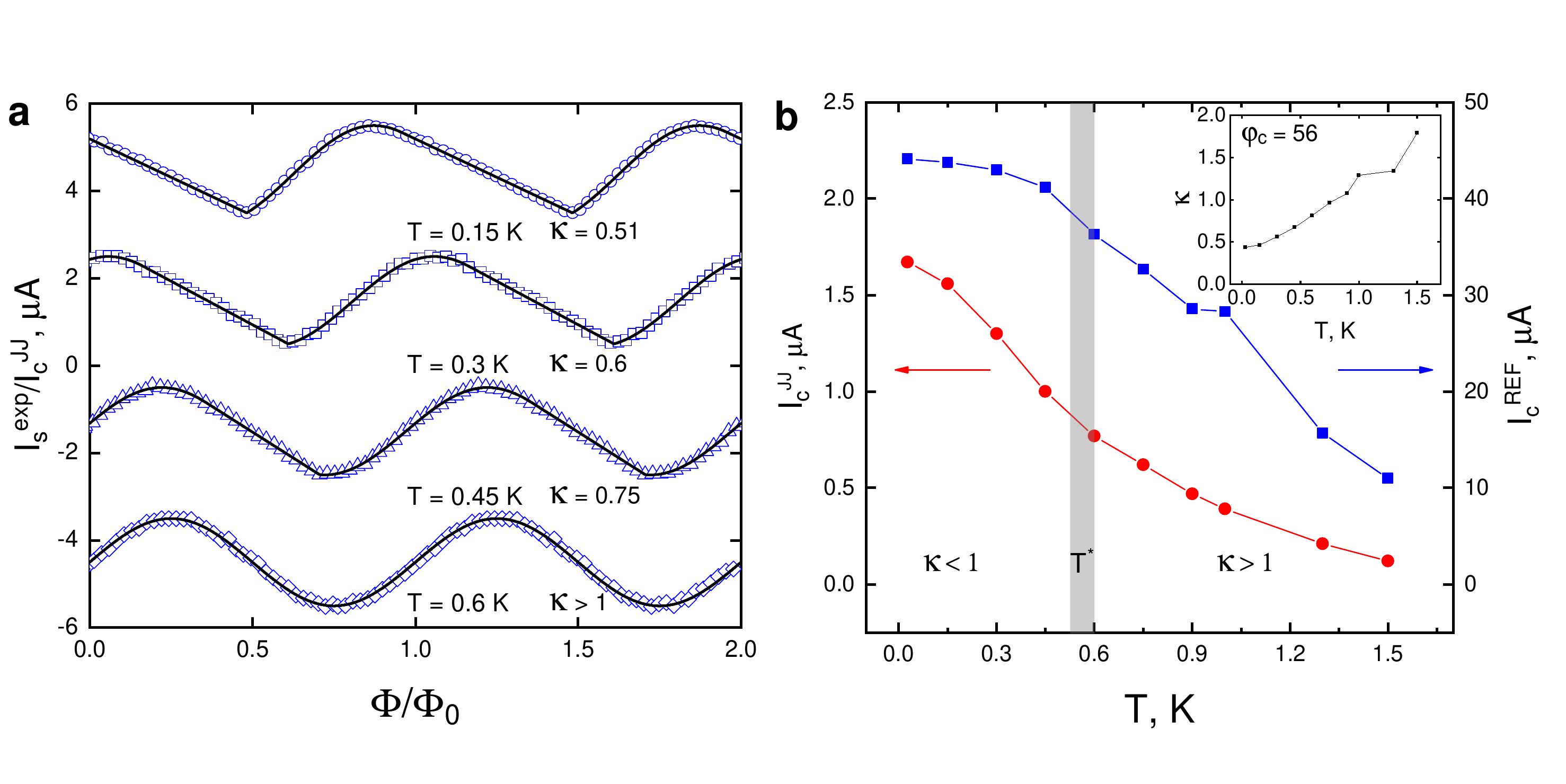} 
			\caption{\textbf{Measurement of ECPR transition with temperature.} (a) ECPRs for different temperatures normalized by the critical currents that were obtained from the fits. The curves are shifted in the vertical direction for clarity. (b) Temperature dependence of the critical current for the studied JJ (red) and for the reference JJ (blue). Parameter $\kappa$ becomes greater than unity in the grey region at a certain $T^*$. Inset graph shows $\kappa = I_c^{REF}(T)/I_c^{JJ}(T)\varphi_c$ where $\varphi_c=56$ was obtained from the fit at T=27 mK. 
			}
			\label{Figure4}
		\end{center}
	\end{figure}

To check the model, we have precisely measured ECPRs near the 0$^{\text{th}}$ and 1$^{\text{st}}$ Fraunhofer maximums (red and blue boxes correspondingly in Figure \ref{Figure1}b). 
In Figure \ref{Figure3}a,b, we show the ECPRs, where each point was obtained by averaging of critical currents from 10,000 I-V measurements. The features in this figure are the same as in Figure \ref{Figure2}b,d.  We fitted the ECPR in Figure \ref{Figure3}a with the model, and the ECPR in Figure \ref{Figure3}b with a sine. Both fits perfectly describe the experiment, therefore we conclude that Figure \ref{Figure3}b shows sinusoidal CPR of the studied JJ that was measured in the "True" regime.

In Figure \ref{Figure3}c the extracted phases are presented for the "False" measurement regime with $\kappa=0.47$\cite{ginzburg2018determination} (see Supplementary Information for details).
Both of the JJs show regions of constant phase, which implies that the role of the reference junction switches between the two JJs in spite of the high asymmetry. 
For this reason, each JJ in the SQUID alternately displays the CPR of the opposite JJ, leading to consecutive linear and sinusoidal segments in the ECPR. The phase dynamics in this case may be seen explicitly in the Supporting Video 1. 

On the other hand, for $\kappa = 1.3$ the phase of the reference JJ is well localized near the critical phase, and the phase of the studied JJ $\delta^*(\Phi)=\varphi_c+2\pi\Phi/\Phi_0$ increases linearly with the applied flux (Figure \ref{Figure3}d). Such phase behavior is an evidence of accurate CPR measurements. 

Parameter $\kappa$ may be controlled not only by the magnetic field but the temperature.
We performed ECPR measurements for the temperatures lower than the critical temperature $T_c^{JJ}\approx 1.7\; K$ of an S-TI-S JJ and observed a gradual transition from skewed oscillations to sinusoidal ones (see Figure \ref{Figure4}a).
Similar results may be interpreted (see, for example, Ref.\cite{kayyalha2020highly}) in the frame of a universal rule, that the CPR should become sinusoidal when the temperature approaches $T_c^{JJ}$ \cite{golubov2004current}. 
However, in our case, it is due to rise of asymmetry $I_c^{REF}/I_c^{JJ}$, which leads to increasing $\kappa$ as shown in the inset of Figure \ref{Figure4}b. 
The observed oscillations appear sinusoidal for the first time at 600 mK, thus $\kappa$ becomes greater than 1 at a certain $T^*$ close to 600 mK (gray region in Figure \ref{Figure4}b).

Notably, the "False" CPR regime makes extraction of the studied JJs critical current as $I_c^{JJ}(H)=I_c^{SQUID}(H)-\langle I_c^{SQUID}(H)\rangle$ misleading, since an average value of the SQUID oscillations does not coincide with $I_c^{REF}$ when $\kappa<1$.

In order to generalize the condition of an accurate CPR measurement for the case of arbitrary SQUID components let us consider the following examples (i)-(iii), where in (i) a linear CPR is used to measure an arbitrary CPR, in (ii) a sinusoidal CPR is used to measure a skewed CPR, and in (iii) a skewed CPR is used to measure a sinusoidal CPR.   

\begin{figure}[h!]
		\begin{center}
			\includegraphics[width=16cm]{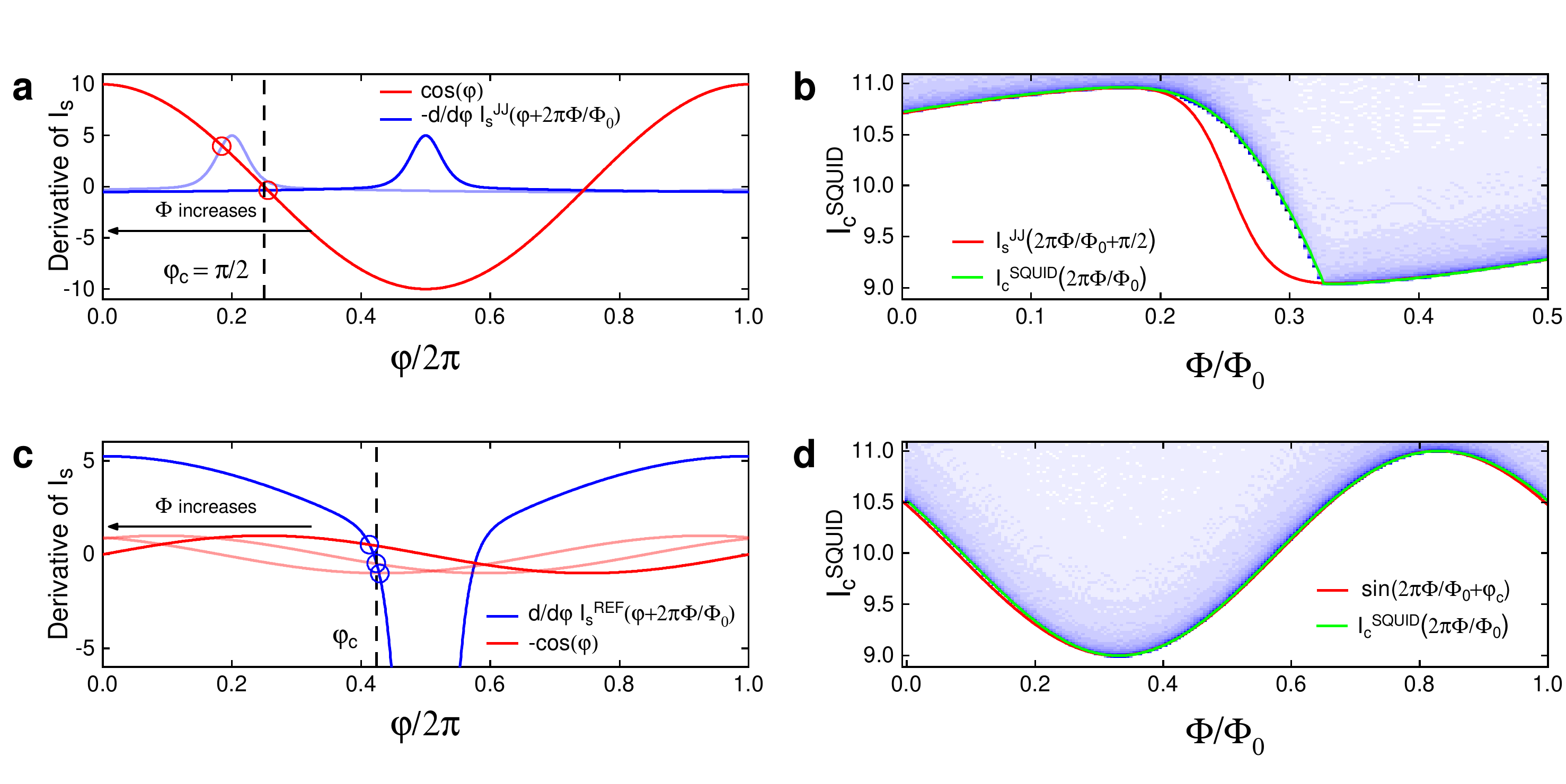} 
		\caption{\textbf{Prediction of measurements with a (non-)skewed reference CPR.} (a) Graphical representation of the equation (\ref{eq4}) for a sinusoidal reference CPR (red) and a skewed measured CPR (blue). Red circles denote solutions $\varphi^*(\Phi)$. (b) The color map is the $dV/dI (\Phi)$ map obtained from the RSJ model and displays the $I_c^{SQUID}(\Phi)$ dependence.The skewed part of the SQUID oscillations does not match with the studied CPR that is denoted by the red curve. (c) Graphical representation of the equation \ref{eq4} for a skewed reference CPR (blue) and a sinusoidal measured CPR (red). Blue circles denote solutions $\varphi^*(\Phi)$. (d) The color map obtained from the RSJ model that displays SQUIDs critical current dependence on the applied magnetic flux. The SQUID oscillations match with the underlying CPR in this case.
			}
			\label{Figure5}
		\end{center}
	\end{figure}

(i) This case generalizes the method of an asymmetric SQUID with a nanobridge that acts as a reference JJ. In Supporting Information, we derive the condition for the "True" measurement regime for an arbitrary studied CPR: $\kappa>\max \partial_{\delta}I_s^{JJ}(\delta)/I_c^{JJ}$. This condition is stricter then $\kappa>1$, and makes it complicated to measure skewed CPRs due to high magnitude of the supercurrents derivative.
If $\kappa$ does not fulfill the condition, there will be both CPRs alternately displayed in the SQUID oscillations, thus the skewed parts will be masked. The described example provides an alternative explanation of the results in Ref. \cite{kayyalha2020highly,murani2017ballistic,li2019zeeman}.

(ii) Let us consider an asymmetric SQUID, where the reference CPR is $I_s^{REF}(\varphi)=I_c^{REF}\sin(\varphi)$ and the studied JJ has a skewed CPR $I_s^{JJ}(\delta)$, which can be observed, for example, in ballistic SNS JJs or point-contact JJs \cite{golubov2004current} (see Supporting Information for details).
In this case the equation (\ref{eq4}) may be graphically displayed as in Figure \ref{Figure5}a.
Due to high skewness of the studied CPR, there is a peak in the derivative (blue line). When the magnetic field is applied, the peak slides, thus the intersection (red circle) with the derivative of the reference CPR (red line) leaves the origin of the critical phase $\varphi_c=\pi/2$ of the sinusoidal CPR, giving rise to non-localized solutions $\varphi^*(\Phi)$. 
This leads to a significant deviation of the SQUID oscillations from the underlying CPR (Figure \ref{Figure5}b), where the most skewed part of the CPR (red line) is masked. The observed feature of both angle-shaped minimums and smooth maximums may be an immediate sign of inaccurate measurement, if observed in the experiment. The angle-shaped minimum originates from a jump-like behaviour of phases as in Figure \ref{Figure3}c. One may explicitly observe the phase dynamics in case of the described example in Supporting Video 2. From the Video 2 and Figure \ref{Figure5}a it is evident that the jump-like phase behaviour arises when the intersection of derivatives jumps to the critical phase after being transferred from its origin by the peak of the studied CPRs derivative. This may explain angle-shaped features in Refs \cite{della2007measurement,lee2015ultimately}, where an SIS reference JJ was used. 

Generally speaking, we can conclude that there is a fundamental limitation of any asymmetric SQUID method. 
If the peak in the studied CPRs derivative is too high, there might be a range of phases $\delta^*(\Phi)$ which do not appear as solutions of (\ref{eq4}). 
Therefore, the corresponding parts of the studied CPR can not be recovered from the data. 
To have the full range of $\delta^*(\Phi)$ available, the following necessary condition should be satisfied:
  \begin{equation}
\begin{cases}
\max \partial_{\varphi}I_s^{REF}>\max-\partial_{\delta}I_s^{JJ}\\
\min \partial_{\varphi}I_s^{REF}<\min-\partial_{\delta}I_s^{JJ}
\label{eq6}
\end{cases},
    \end{equation}
which says that the oscillations of the studied CPR derivative should fall inside the range of reference CPRs derivative. The described problem is evident in Figure \ref{Figure2}c and Figure \ref{Figure3}c. In this case the condition (\ref{eq6}) is not satisfied, thus there are approximately 50\% of $\delta^*(\Phi)$ lost. 

The condition (\ref{eq6}) is necessary but not sufficient. In example (ii), this condition is satisfied, however the SQUID oscillations still do not coincide with the CPR. 
The phase error $|\varphi_c-\varphi^*(\Phi)|$ is limited by the $\varphi-$coordinate of the intersection of the derivative peaks maximum and the reference CPRs derivative (see Figure \ref{Figure5}a). 
Therefore, the error is mediated by the asymmetry of the magnitudes of the derivatives, increasing which ensures an accurate CPR measurement. 

(iii) In this example, the skewed CPR from (ii) is used as a reference to measure a sinusoidal CPR, thus a high asymmetry in derivatives of the CPRs is established. In Figure \ref{Figure5}c, the peak of the reference CPR "scans" the derivative of the studied CPR. 
Therefore, the solutions $\varphi^*(\Phi)$ are localized near the critical phase, enabling accurate CPR measurements as shown in Figure \ref{Figure5}d.

Thus, we consider a point contact or a ballistic SNS with a highly-skewed CPR to be a good choice for a reference JJ. They allow measurement of mildly non-sinusoidal CPRs. 
However, measurements of highly skewed CPRs may be complicated due to the high magnitude of the studied CPRs derivative. 
The best candidate for the reference JJ for an asymmetric dc-SQUID method should have a linear CPR (sawtooth shaped) with a low critical phase \cite{pop2010measurement}. In this case the model for an asymmetric SQUID with a nanobridge is applicable, but $\kappa=I_c^{REF}/I_c^{SQUID}\varphi_c$ is much larger due to low critical phase.
	
In summary, we showed that an asymmetric dc-SQUID method works only when high asymmetry in magnitudes of derivatives of the supercurrents is established. The inaccurately measured CPR may be recognized by angle-shaped extremums (arbitrary case) or linear segments (case of a reference nanobridge junction). These features are artifacts of the method and were previously misinterpreted as parts of the real CPR. The condition for an accurate CPR measurement becomes stricter with increasing skewness of the studied CPR. In view of the above, we provided considerations for an accurate CPR measurement that might encourage future experiments with reference CPRs different from those that were used previously.

\section*{Acknowledgements}
It is a pleasure to thank A. A. Golubov, M. Y. Kupriyanov, and S. Kozlov for valuable discussions, and D. Kalashnikov for the help with the device fabrication. This work was supported by the RSF 21-72-00140. E-beam lithography is supported by the Ministry of Science and Higher Education of the Russian Federation (No. FSMG-2021-0005) and was done in the MIPT Shared Facilities Center.

\section*{Author contributions statement}
 I.B. did the theoretical analysis and I$_c$(H), I$_c$(T) fitting, A.K. fabricated the samples, I.B., A.K. and D.B. conducted the measurements, V.S.S. supervised the project, I.B. wrote the manuscript with the contribution from all authors.

\section*{Additional information}
The
authors declare no competing interests.

\providecommand{\latin}[1]{#1}
\makeatletter
\providecommand{\doi}
  {\begingroup\let\do\@makeother\dospecials
  \catcode`\{=1 \catcode`\}=2 \doi@aux}
\providecommand{\doi@aux}[1]{\endgroup\texttt{#1}}
\makeatother
\providecommand*\mcitethebibliography{\thebibliography}
\csname @ifundefined\endcsname{endmcitethebibliography}
  {\let\endmcitethebibliography\endthebibliography}{}


\begin{mcitethebibliography}{22}
\providecommand*\natexlab[1]{#1}
\providecommand*\mciteSetBstSublistMode[1]{}
\providecommand*\mciteSetBstMaxWidthForm[2]{}
\providecommand*\mciteBstWouldAddEndPuncttrue
  {\def\EndOfBibitem{\unskip.}}
\providecommand*\mciteBstWouldAddEndPunctfalse
  {\let\EndOfBibitem\relax}
\providecommand*\mciteSetBstMidEndSepPunct[3]{}
\providecommand*\mciteSetBstSublistLabelBeginEnd[3]{}
\providecommand*\EndOfBibitem{}
\mciteSetBstSublistMode{f}
\mciteSetBstMaxWidthForm{subitem}{(\alph{mcitesubitemcount})}
\mciteSetBstSublistLabelBeginEnd
  {\mcitemaxwidthsubitemform\space}
  {\relax}
  {\relax}

\bibitem[Troeman \latin{et~al.}(2008)Troeman, Van Der~Ploeg, Il’Ichev, Meyer,
  Golubov, Kupriyanov, and Hilgenkamp]{troeman2008temperature}
Troeman,~A.; Van Der~Ploeg,~S.; Il’Ichev,~E.; Meyer,~H.-G.; Golubov,~A.~A.;
  Kupriyanov,~M.~Y.; Hilgenkamp,~H. Temperature dependence measurements of the
  supercurrent-phase relationship in niobium nanobridges. \emph{Physical Review
  B} \textbf{2008}, \emph{77}, 024509\relax
\mciteBstWouldAddEndPuncttrue
\mciteSetBstMidEndSepPunct{\mcitedefaultmidpunct}
{\mcitedefaultendpunct}{\mcitedefaultseppunct}\relax
\EndOfBibitem
\bibitem[Spanton \latin{et~al.}(2017)Spanton, Deng, Vaitiek{\.e}nas, Krogstrup,
  Nyg{\aa}rd, Marcus, and Moler]{spanton2017current}
Spanton,~E.~M.; Deng,~M.; Vaitiek{\.e}nas,~S.; Krogstrup,~P.; Nyg{\aa}rd,~J.;
  Marcus,~C.~M.; Moler,~K.~A. Current--phase relations of few-mode InAs
  nanowire Josephson junctions. \emph{Nature Physics} \textbf{2017}, \emph{13},
  1177--1181\relax
\mciteBstWouldAddEndPuncttrue
\mciteSetBstMidEndSepPunct{\mcitedefaultmidpunct}
{\mcitedefaultendpunct}{\mcitedefaultseppunct}\relax
\EndOfBibitem
\bibitem[Hart \latin{et~al.}(2019)Hart, Cui, M{\'e}nard, Deng, Antipov,
  Lutchyn, Krogstrup, Marcus, and Moler]{hart2019current}
Hart,~S.; Cui,~Z.; M{\'e}nard,~G.; Deng,~M.; Antipov,~A.~E.; Lutchyn,~R.~M.;
  Krogstrup,~P.; Marcus,~C.~M.; Moler,~K.~A. Current-phase relations of InAs
  nanowire Josephson junctions: From interacting to multimode regimes.
  \emph{Physical Review B} \textbf{2019}, \emph{100}, 064523\relax
\mciteBstWouldAddEndPuncttrue
\mciteSetBstMidEndSepPunct{\mcitedefaultmidpunct}
{\mcitedefaultendpunct}{\mcitedefaultseppunct}\relax
\EndOfBibitem
\bibitem[Frolov \latin{et~al.}(2004)Frolov, Van~Harlingen, Oboznov, Bolginov,
  and Ryazanov]{frolov2004measurement}
Frolov,~S.; Van~Harlingen,~D.; Oboznov,~V.; Bolginov,~V.; Ryazanov,~V.
  Measurement of the current-phase relation of
  superconductor/ferromagnet/superconductor $\pi$ Josephson junctions.
  \emph{Physical Review B} \textbf{2004}, \emph{70}, 144505\relax
\mciteBstWouldAddEndPuncttrue
\mciteSetBstMidEndSepPunct{\mcitedefaultmidpunct}
{\mcitedefaultendpunct}{\mcitedefaultseppunct}\relax
\EndOfBibitem
\bibitem[Sochnikov \latin{et~al.}(2013)Sochnikov, Bestwick, Williams, Lippman,
  Fisher, Goldhaber-Gordon, Kirtley, and Moler]{sochnikov2013direct}
Sochnikov,~I.; Bestwick,~A.~J.; Williams,~J.~R.; Lippman,~T.~M.; Fisher,~I.~R.;
  Goldhaber-Gordon,~D.; Kirtley,~J.~R.; Moler,~K.~A. Direct measurement of
  current-phase relations in superconductor/topological
  insulator/superconductor junctions. \emph{Nano letters} \textbf{2013},
  \emph{13}, 3086--3092\relax
\mciteBstWouldAddEndPuncttrue
\mciteSetBstMidEndSepPunct{\mcitedefaultmidpunct}
{\mcitedefaultendpunct}{\mcitedefaultseppunct}\relax
\EndOfBibitem
\bibitem[Sochnikov \latin{et~al.}(2015)Sochnikov, Maier, Watson, Kirtley,
  Gould, Tkachov, Hankiewicz, Br{\"u}ne, Buhmann, Molenkamp, \latin{et~al.}
  others]{sochnikov2015nonsinusoidal}
Sochnikov,~I.; Maier,~L.; Watson,~C.~A.; Kirtley,~J.~R.; Gould,~C.;
  Tkachov,~G.; Hankiewicz,~E.~M.; Br{\"u}ne,~C.; Buhmann,~H.; Molenkamp,~L.~W.,
  \latin{et~al.}  Nonsinusoidal current-phase relationship in Josephson
  junctions from the 3D topological insulator HgTe. \emph{Physical review
  letters} \textbf{2015}, \emph{114}, 066801\relax
\mciteBstWouldAddEndPuncttrue
\mciteSetBstMidEndSepPunct{\mcitedefaultmidpunct}
{\mcitedefaultendpunct}{\mcitedefaultseppunct}\relax
\EndOfBibitem
\bibitem[Golubov \latin{et~al.}(2004)Golubov, Kupriyanov, and
  Il’Ichev]{golubov2004current}
Golubov,~A.~A.; Kupriyanov,~M.~Y.; Il’Ichev,~E. The current-phase relation in
  Josephson junctions. \emph{Reviews of modern physics} \textbf{2004},
  \emph{76}, 411\relax
\mciteBstWouldAddEndPuncttrue
\mciteSetBstMidEndSepPunct{\mcitedefaultmidpunct}
{\mcitedefaultendpunct}{\mcitedefaultseppunct}\relax
\EndOfBibitem
\bibitem[Haller \latin{et~al.}(2022)Haller, F{\"u}l{\"o}p, Indolese, Ridderbos,
  Kraft, Cheung, Ungerer, Watanabe, Taniguchi, Beckmann, \latin{et~al.}
  others]{haller2022phase}
Haller,~R.; F{\"u}l{\"o}p,~G.; Indolese,~D.; Ridderbos,~J.; Kraft,~R.;
  Cheung,~L.~Y.; Ungerer,~J.~H.; Watanabe,~K.; Taniguchi,~T.; Beckmann,~D.,
  \latin{et~al.}  Phase-dependent microwave response of a graphene Josephson
  junction. \emph{Physical Review Research} \textbf{2022}, \emph{4},
  013198\relax
\mciteBstWouldAddEndPuncttrue
\mciteSetBstMidEndSepPunct{\mcitedefaultmidpunct}
{\mcitedefaultendpunct}{\mcitedefaultseppunct}\relax
\EndOfBibitem
\bibitem[Kayyalha \latin{et~al.}(2020)Kayyalha, Kazakov, Miotkowski,
  Khlebnikov, Rokhinson, and Chen]{kayyalha2020highly}
Kayyalha,~M.; Kazakov,~A.; Miotkowski,~I.; Khlebnikov,~S.; Rokhinson,~L.~P.;
  Chen,~Y.~P. Highly skewed current--phase relation in
  superconductor--topological insulator--superconductor Josephson junctions.
  \emph{npj Quantum Materials} \textbf{2020}, \emph{5}, 1--7\relax
\mciteBstWouldAddEndPuncttrue
\mciteSetBstMidEndSepPunct{\mcitedefaultmidpunct}
{\mcitedefaultendpunct}{\mcitedefaultseppunct}\relax
\EndOfBibitem
\bibitem[Li \latin{et~al.}(2019)Li, De~Ronde, De~Boer, Ridderbos, Zwanenburg,
  Huang, Golubov, and Brinkman]{li2019zeeman}
Li,~C.; De~Ronde,~B.; De~Boer,~J.; Ridderbos,~J.; Zwanenburg,~F.; Huang,~Y.;
  Golubov,~A.; Brinkman,~A. Zeeman-effect-induced 0- $\pi$ transitions in
  ballistic Dirac semimetal Josephson junctions. \emph{Physical review letters}
  \textbf{2019}, \emph{123}, 026802\relax
\mciteBstWouldAddEndPuncttrue
\mciteSetBstMidEndSepPunct{\mcitedefaultmidpunct}
{\mcitedefaultendpunct}{\mcitedefaultseppunct}\relax
\EndOfBibitem
\bibitem[Kayyalha \latin{et~al.}(2019)Kayyalha, Kargarian, Kazakov, Miotkowski,
  Galitski, Yakovenko, Rokhinson, and Chen]{kayyalha2019anomalous}
Kayyalha,~M.; Kargarian,~M.; Kazakov,~A.; Miotkowski,~I.; Galitski,~V.~M.;
  Yakovenko,~V.~M.; Rokhinson,~L.~P.; Chen,~Y.~P. Anomalous low-temperature
  enhancement of supercurrent in topological-insulator nanoribbon Josephson
  junctions: Evidence for low-energy Andreev bound states. \emph{Physical
  Review Letters} \textbf{2019}, \emph{122}, 047003\relax
\mciteBstWouldAddEndPuncttrue
\mciteSetBstMidEndSepPunct{\mcitedefaultmidpunct}
{\mcitedefaultendpunct}{\mcitedefaultseppunct}\relax
\EndOfBibitem
\bibitem[Nichele \latin{et~al.}(2020)Nichele, Portol{\'e}s, Fornieri, Whiticar,
  Drachmann, Gronin, Wang, Gardner, Thomas, Hatke, \latin{et~al.}
  others]{nichele2020relating}
Nichele,~F.; Portol{\'e}s,~E.; Fornieri,~A.; Whiticar,~A.~M.; Drachmann,~A.~C.;
  Gronin,~S.; Wang,~T.; Gardner,~G.; Thomas,~C.; Hatke,~A., \latin{et~al.}
  Relating Andreev bound states and supercurrents in hybrid Josephson
  junctions. \emph{Physical Review Letters} \textbf{2020}, \emph{124},
  226801\relax
\mciteBstWouldAddEndPuncttrue
\mciteSetBstMidEndSepPunct{\mcitedefaultmidpunct}
{\mcitedefaultendpunct}{\mcitedefaultseppunct}\relax
\EndOfBibitem
\bibitem[Nanda \latin{et~al.}(2017)Nanda, Aguilera-Servin, Rakyta,
  Korm{\'a}nyos, Kleiner, Koelle, Watanabe, Taniguchi, Vandersypen, and
  Goswami]{nanda2017current}
Nanda,~G.; Aguilera-Servin,~J.~L.; Rakyta,~P.; Korm{\'a}nyos,~A.; Kleiner,~R.;
  Koelle,~D.; Watanabe,~K.; Taniguchi,~T.; Vandersypen,~L.~M.; Goswami,~S.
  Current-phase relation of ballistic graphene Josephson junctions. \emph{Nano
  Letters} \textbf{2017}, \emph{17}, 3396--3401\relax
\mciteBstWouldAddEndPuncttrue
\mciteSetBstMidEndSepPunct{\mcitedefaultmidpunct}
{\mcitedefaultendpunct}{\mcitedefaultseppunct}\relax
\EndOfBibitem
\bibitem[Della~Rocca \latin{et~al.}(2007)Della~Rocca, Chauvin, Huard, Pothier,
  Esteve, and Urbina]{della2007measurement}
Della~Rocca,~M.; Chauvin,~M.; Huard,~B.; Pothier,~H.; Esteve,~D.; Urbina,~C.
  Measurement of the current-phase relation of superconducting atomic contacts.
  \emph{Physical review letters} \textbf{2007}, \emph{99}, 127005\relax
\mciteBstWouldAddEndPuncttrue
\mciteSetBstMidEndSepPunct{\mcitedefaultmidpunct}
{\mcitedefaultendpunct}{\mcitedefaultseppunct}\relax
\EndOfBibitem
\bibitem[Lee \latin{et~al.}(2015)Lee, Kim, Jhi, and Lee]{lee2015ultimately}
Lee,~G.-H.; Kim,~S.; Jhi,~S.-H.; Lee,~H.-J. Ultimately short ballistic vertical
  graphene Josephson junctions. \emph{Nature communications} \textbf{2015},
  \emph{6}, 1--9\relax
\mciteBstWouldAddEndPuncttrue
\mciteSetBstMidEndSepPunct{\mcitedefaultmidpunct}
{\mcitedefaultendpunct}{\mcitedefaultseppunct}\relax
\EndOfBibitem
\bibitem[Murani \latin{et~al.}(2017)Murani, Kasumov, Sengupta, Kasumov, Volkov,
  Khodos, Brisset, Delagrange, Chepelianskii, Deblock, \latin{et~al.}
  others]{murani2017ballistic}
Murani,~A.; Kasumov,~A.; Sengupta,~S.; Kasumov,~Y.~A.; Volkov,~V.; Khodos,~I.;
  Brisset,~F.; Delagrange,~R.; Chepelianskii,~A.; Deblock,~R., \latin{et~al.}
  Ballistic edge states in Bismuth nanowires revealed by SQUID interferometry.
  \emph{Nature Communications} \textbf{2017}, \emph{8}, 1--7\relax
\mciteBstWouldAddEndPuncttrue
\mciteSetBstMidEndSepPunct{\mcitedefaultmidpunct}
{\mcitedefaultendpunct}{\mcitedefaultseppunct}\relax
\EndOfBibitem
\bibitem[Ginzburg \latin{et~al.}(2018)Ginzburg, Batov, Bol’ginov, Egorov,
  Chichkov, Shchegolev, Klenov, Soloviev, Bakurskiy, and
  Kupriyanov]{ginzburg2018determination}
Ginzburg,~L.~V.; Batov,~I.; Bol’ginov,~V.; Egorov,~S.~V.; Chichkov,~V.;
  Shchegolev,~A.~E.; Klenov,~N.~V.; Soloviev,~I.; Bakurskiy,~S.~V.;
  Kupriyanov,~M.~Y. Determination of the current--phase relation in Josephson
  junctions by means of an asymmetric two-junction SQUID. \emph{JETP Letters}
  \textbf{2018}, \emph{107}, 48--54\relax
\mciteBstWouldAddEndPuncttrue
\mciteSetBstMidEndSepPunct{\mcitedefaultmidpunct}
{\mcitedefaultendpunct}{\mcitedefaultseppunct}\relax
\EndOfBibitem
\bibitem[Murphy and Bezryadin(2017)Murphy, and Bezryadin]{murphy2017asymmetric}
Murphy,~A.; Bezryadin,~A. Asymmetric nanowire SQUID: Linear current-phase
  relation, stochastic switching, and symmetries. \emph{Physical Review B}
  \textbf{2017}, \emph{96}, 094507\relax
\mciteBstWouldAddEndPuncttrue
\mciteSetBstMidEndSepPunct{\mcitedefaultmidpunct}
{\mcitedefaultendpunct}{\mcitedefaultseppunct}\relax
\EndOfBibitem
\bibitem[Vijay \latin{et~al.}(2009)Vijay, Sau, Cohen, and
  Siddiqi]{vijay2009optimizing}
Vijay,~R.; Sau,~J.; Cohen,~M.~L.; Siddiqi,~I. Optimizing anharmonicity in
  nanoscale weak link Josephson junction oscillators. \emph{Physical review
  letters} \textbf{2009}, \emph{103}, 087003\relax
\mciteBstWouldAddEndPuncttrue
\mciteSetBstMidEndSepPunct{\mcitedefaultmidpunct}
{\mcitedefaultendpunct}{\mcitedefaultseppunct}\relax
\EndOfBibitem
\bibitem[Dausy \latin{et~al.}(2021)Dausy, Nulens, Raes, Van~Bael, and Van~de
  Vondel]{dausy2021impact}
Dausy,~H.; Nulens,~L.; Raes,~B.; Van~Bael,~M.~J.; Van~de Vondel,~J. Impact of
  kinetic inductance on the critical-current oscillations of nanobridge SQUIDs.
  \emph{Physical Review Applied} \textbf{2021}, \emph{16}, 024013\relax
\mciteBstWouldAddEndPuncttrue
\mciteSetBstMidEndSepPunct{\mcitedefaultmidpunct}
{\mcitedefaultendpunct}{\mcitedefaultseppunct}\relax
\EndOfBibitem
\bibitem[Pop \latin{et~al.}(2010)Pop, Protopopov, Lecocq, Peng, Pannetier,
  Buisson, and Guichard]{pop2010measurement}
Pop,~I.~M.; Protopopov,~I.; Lecocq,~F.; Peng,~Z.; Pannetier,~B.; Buisson,~O.;
  Guichard,~W. Measurement of the effect of quantum phase slips in a Josephson
  junction chain. \emph{Nature Physics} \textbf{2010}, \emph{6}, 589--592\relax
\mciteBstWouldAddEndPuncttrue
\mciteSetBstMidEndSepPunct{\mcitedefaultmidpunct}
{\mcitedefaultendpunct}{\mcitedefaultseppunct}\relax
\EndOfBibitem
\end{mcitethebibliography}
\end{document}


\begin{center}
		\Large
		Supporting Information
		
		Limitations of the current-phase relation measurements by an asymmetric dc-SQUID
	\end{center}

\subsection*{Transport measurements}
We used a dilution refrigerator BlueFors LD-250, equipped with a 9T superconducting solenoid.
All the measurements were performed at the temperature of $27\;mK$ if not stated otherwise.
Yokogawa GS-200 was used to charge the solenoid. 
In order to measure I-V characteristic we applied a 10 $\text{V}_{\text{pp}}$ sinusoidal signal with frequancy $f=78.688$ Hz through a 50 kOhm resistor by Low-distortion Function Generator SRS DS360. 
The voltage drop on the sample was amplified by the low noise preamplifier SRS SR560 and then measured by NI-9234 ADC. 
The second channel of ADC measured the voltage, provided by the generator, i.e., the current through the sample. 
At each value of the magnetic field, we measured 10,000 I-V characteristics, determined the critical current by the threshold method, and averaged the list of the obtained values to get a single point in the $I_c^{SQUID}(H)$ curve. 
The value of the critical current weakly depends on the threshold, since the I-V characteristic is very sharp at $I=I_c$.
QFilter from QDevil was placed at the MC plate of the cryostat.

\subsection*{Device fabrication}
We have developed a two-step process for the device fabrication.
On the first step, the grid of nanobridges or symmetric SQUIDs (devices that consist of two identical nanobridges, see Figure S4) was formed via an electron beam lithography on MMA/PMMA bilayer resist, followed by e-beam evaporation of 20 nm of Nb and a lift-off. Then the flakes of Bi$_2$Te$_2$Se crystals (grown by a Bridgman method) were mechanically exfoliated and transferred to the substrate with nanobridges using blue tape.
The flakes with an appropriate size and location were selected for the S-TI-S JJs. 
On the second step, another electron beam lithography on PMMA resist was done in order to define Josephson junctions, SQUID loops, and contacts. 
The substrates were exposed to the 20 sec of Ar-plasma etching process prior to the deposition of a 100 nm thin Nb film by magnetron sputtering that was followed by  lift-off. 
This two-step technology was developed for two reasons. 
(i) The thickness of the flakes is around 50-70 nm, therefore, in order to have good electrical contact between Nb and the topological insulator, the thickness of the deposited metal film should be greater than that of the flake. 
At the same time, the thickness of the nanobridges should be small enough in order to have a smaller weak link cross-section for appropriate critical currents. 
(ii) It is necessary to avoid heating the flakes during the e-beam evaporation of the nanobridges.

We would like to note, that the properties of nanobridges (critical current and critical phase) are reproducible from device to device if fabricated in the same process, but are not reproducible between different processes. 
See Figures S5 and S6 for details.  

\subsection*{Model details}
Let us consider an asymmetric SQUID that consists of the studied JJ and a reference JJ that have CPRs $I_s^{JJ}(\delta)$ and $I_s^{REF}(\varphi)$ correspondingly. 
Phase differences between the superconducting leads of the JJs are related by a well-known formula $\delta-\varphi = 2\pi (\Phi-L^{SQUID} I_{circ})/\Phi_0\approx 2\pi \Phi/\Phi_0$, where the contribution of a SQUID loop inductance  $L^{SQUID}$ is typically neglected for similar to our case geometric dimensions (see, for example, methods section in Ref. \cite{3}).

The critical current of the SQUID is a maximum non-dissipative current that can be present in the SQUID:
			\begin{equation}
			I_c^{SQUID}(\Phi)=\max\limits_{\varphi}\left[I_s^{REF}(\varphi)+I_s^{JJ}(\varphi+2\pi\Phi/\Phi_0 ) \right].
		\end{equation}
The maximum is reached at a certain phase difference $\varphi=\varphi^*(\Phi)$, which is impossible to determine without knowing the exact CPR functions.
Further development is usually made under the assumption of a high asymmetry $I_c^{REF}\gg I_c^{JJ}$, which allows to approximately write $\varphi^*(\Phi)\approx\varphi_c$, where $\varphi_c$ is the critical phase difference at which the critical current of the reference JJ is reached. If $\varphi^*(\Phi)$ is well localized near $\varphi_c$, the supercurrent through the reference JJ does not change with the flux.
Therefore, the supercurrent of the studied JJ may be approximated as
 	\begin{equation}
	I_{s}^{JJ}(\varphi^*+2\pi\Phi/\Phi_0 ) \approx	I_{s}^{exp}(\varphi_c+2\pi\Phi/\Phi_0 ) = I_c^{SQUID}(\Phi)-I_c^{REF}.
	\label{eq3}
	\end{equation}
Here we introduced $I_s^{exp}(\Phi)$ that is expected to be the studied current-phase relation (ECPR). 
It is usually assumed \cite{4}, that $I_c^{REF} = \langle I_c^{SQUID}(\Phi)\rangle $ to represent (\ref{eq3}) in measurable quantities. The brackets denote averaging over one period of oscillations. 

The equation to determine the maximizing phase difference $\varphi^*(\Phi)$ is
\begin{equation}
     \partial_{\varphi}I_s^{REF}(\varphi) +\partial_{\varphi} I_s^{JJ}(\varphi+2\pi\Phi/\Phi_0) = 0,
        \label{eq4}
\end{equation}

We solve the following system of equations to calculate the ECPR for the SQUID with JJs that have sinusoidal and linear CPRs:
\begin{equation}
		\begin{cases}
			I_s^{REF}(\varphi) = I_c^{REF}\frac{(\varphi-\varphi_c\mod{2\pi})+\varphi_c-2\pi}{\varphi_c},\\
			I_s^{JJ}(\delta)=I_c^{JJ}\sin\delta,\\
		   I_s^{SQUID}( \varphi,\delta) = I_s^{REF}(\varphi)+I_s^{JJ}(\delta ),\\	I_c^{SQUID}(\Phi)=\max\limits_{\varphi}\left[I_s^{SQUID}(\varphi,\delta) \right],\\
			\delta-\varphi = 2\pi\Phi/\Phi_0,\\
			I_{s}^{exp}(\Phi) = I_c^{SQUID}(\Phi)-I_c^{REF}.
		\end{cases}
	\end{equation}
The critical current for the linear CPR is reached at $\varphi_c\pm2\pi k$. We would like to note, that such a sawtooth-shaped CPR is an upper part of a multivalued CPR of the nanobridge. The phases of the JJs are defined up to a $2\pi k$, thus we will consider $\varphi^*(\Phi)$ solutions that fall in the $(\varphi_c-2\pi,\varphi_c)$ range. 

	The analytical solution is the following:
	\begin{equation}
	        \varphi^*(\Phi) =
	       \begin{cases} -2\pi\Phi/\Phi_0+\arccos\left({-\frac{I_{c}^{REF}}{I_c^{JJ}\varphi_c}}\right)+2\pi \operatorname{floor}(\Phi/\Phi_0)+2\pi k, \text{if}\\
	       I_s^{SQUID}(\varphi^*(\Phi),\varphi^*(\Phi)+2\pi\Phi/\Phi_0)\ge I_s^{SQUID}(\varphi_c,\varphi_c+2\pi\Phi/\Phi_0)\; \text{and}\\
	       \varphi^*(\Phi)\in(\varphi_c-2\pi,\varphi_c);\\
	       \varphi_c,\; \text{otherwise}.
\end{cases}
\label{eqs1}
	\end{equation}

The first expression follows from the condition on $\varphi^*$:
\begin{equation}
\kappa+\cos(\varphi+2\pi\Phi/\Phi_0)=0,
\end{equation}
gathered with the requirement that the phase over the reference JJ should be bounded. The latter is taken into account by $2\pi \operatorname{floor}(\Phi/\Phi_0)$ term that describes the phase slips when $\varphi^*$ exceeds the critical phase. The First expression in equation (\ref{eqs1}) is a solution if it provides the current in the SQUID larger then at the critical phase (Figure S1a) and falls in the range ($\varphi_c-2\pi,\varphi_c$). 
In other cases $\varphi^*=\varphi_c$. The phase from equation (\ref{eqs1}) leads to the SQUID oscillations that are shown in Figure S1b for different $\kappa$ parameters.

Below we show that in case of an arbitrary studied JJ one will observe alternately linear segments of the reference CPR and parts of the studied CPR.

	\begin{equation}
		\begin{gathered}
			\kappa + \partial_{\varphi}I_s^{JJ}(\varphi^*+2\pi\Phi/\Phi_0)=0,\\
			\varphi^* = -2\pi\Phi/\Phi_0 + (\partial_{\varphi}I_s^{JJ})^{-1}[-\kappa],\\
			I_c^{SQUID}(\Phi)= -2\pi\Phi/\Phi_0 \cdot I_c^{REF}/\varphi_c + I_s^{JJ}((\partial_{\varphi}I_s^{JJ})^{-1}[-\kappa]).
			\label{72}
		\end{gathered}
	\end{equation}
	
The condition for an accurate CPR measurement depends on the skewness of the CPR, i.e, the derivative $\kappa$ of the reference CPR should be always higher then that of the studied CPR: 
$\kappa>\max\partial_{\delta} I_s^{JJ}(\delta)$.

\subsection*{Extraction of the phase differences and CPR of the nanobridge}

It is possible to extract the studied CPR from the data if the CPR of the reference JJ is known and if all phases $\delta^*(\Phi)$ are available. This was shown theoretically and experimentally \cite{6} for a sinusoidal reference CPR. In case if $I_s^{REF}(\varphi)=I_c^{REF}\sin\varphi$ the $\varphi^*(\Phi)$ expression reads
\begin{equation}
\varphi^*(\Phi) = \arccos\left(-\frac{\Phi_0}{2\pi I_c^{REF}}\frac{dI_c^{SQUID}(\Phi)}{d\Phi}\right)+2\pi k.
\end{equation}
Where we used $\delta=\varphi+2\pi\Phi/\Phi_0$ relation.

Let us derive the phase $\varphi^*(\Phi)$ following the same method as in Ref. \cite{6} for $I_s^{JJ}(\delta)=I_c^{JJ}\sin\delta$ :
\begin{equation}
\begin{gathered}
\frac{dI_s^{SQUID}}{d\varphi} = \frac{dI_s^{REF}(\varphi)}{d\varphi}\bigg|_{\varphi=\varphi^*}+I_c^{JJ}\cos(\varphi^*+2\pi\Phi/\Phi_0) =0;               \\
\frac{dI_c^{SQUID}}{d\Phi} = \frac{dI_s^{REF}(\varphi^*)}{d\varphi^*}\frac{d\varphi^*}{d\Phi}+I_c^{JJ}\cos(\varphi^*+2\pi\Phi/\Phi_0)\cdot\left(\frac{2\pi}{\Phi_0}+\frac{d\varphi^*}{d\Phi}\right) =              \\
=\frac{2\pi I_c^{JJ}}{\Phi_0}\cos(\varphi^*+2\pi\Phi/\Phi_0);\\
    \varphi^*(\Phi) = \arccos\left(\frac{\Phi_0}{2\pi I_c^{JJ}}\frac{dI_c^{SQUID}(\Phi)}{d\Phi}\right)-2\pi\Phi/\Phi_0+2\pi k(\Phi).
    \end{gathered}
    \label{eqs2}
\end{equation}

Note that we did not use the condition of high asymmetry that determines which JJ will act as a reference. In fact, this condition is hidden in the $2\pi k(\Phi)$ term. In the experiment, the phase of the studied CPR should revolve and grow with the applied field, thus the phase of the reference JJ should be bounded. In expression for $\varphi^*(\Phi)$, the term $-2\pi\Phi/\Phi_0$ is compensated by $+2\pi k(\Phi)$. In our system, $2\pi k(\Phi)$ accounts for the phase slips, this expression is the $2\pi \operatorname{floor}(\Phi/\Phi_0)$ term in equation (\ref{eqs1}). 

Derivation (\ref{eqs2}) may be applied to any known reference CPR. However, it is challenging to control CPR of the reference JJ in the experiment if there are too many parameters. A sinusoidal CPR does not lead to such problems, however measurement of highly skewed CPR remains complicated due to the reasons discussed in the main text.

In our experiment, we observed sinusoidal CPR of the S-TI-S JJ, which allowed to extract all phases of the JJs in the SQUID and a fragment of the nanobridges' CPR (Figure S2a,b):
\begin{equation}
I_s^{REF}(\varphi^*(\Phi)) = I_c^{SQUID}(\Phi) - I_c^{JJ}\sin(\varphi^*(\Phi)).
\end{equation}
 We observed approximately a half of the nanobridges CPRs, because the range of available phase $\varphi^*$ was less then $2\pi$. The obtained CPR is linear and shows critical phase of about 56 radians. Semi-transparent points in the figure originate from the error of the numerical derivative.

\subsection*{Data processing}
In Figure 1b $dV/dI(\Phi)$ color map for the single S-TI-S JJ (see Figure S3) was shifted and stretch to match the envelope of an asymmetric SQUID oscillations. That was done to see whether the measured ECPR mirrors the S-TI-S properties.

In Figure 4a, we duplicated a single period of oscillations to present two periods in all the figures and to show angle-shaped minimums clearly.

For simulations presented in Figure 5 we used a non-sinusoidal CPR from the KO-2 model \cite{5}:
\begin{equation}
I_s^{JJ}(\delta) = \#\sin(\delta/2)\tanh\frac{\Delta\cos(\delta/2)}{2T}.
\label{eqsns}
\end{equation}
Here $\Delta/2T$ controls the skewness of the CPR and was set to be 10 in all computations. Asymmetry in critical currents $I_c^{JJ}/I_c^{REF}$ was set to be 1/10.

\subsection*{Supporting video 1}
Supporting video 1 consists of three animated graphs (i)-(iii) and describes an asymmetric SQUID that consists of two JJs with sinusoidal and linear CPRs. In the graphs, the applied magnetic flux $\Phi(t)$ is parameterized with time.

Graph (i) shows the solution $\delta^*(\Phi)$ of equation 
\begin{equation}
     \kappa\cdot(1-\operatorname{\delta}(\varphi^*(\Phi)-\varphi_c)) =-\cos(\delta^*(\Phi)),
     \label{sa1}
\end{equation}
that maximizes the SQUIDs critical current. Parameter $\kappa=0.5$, $\delta(\cdot)$ is a Dirac delta-function. The solution is the intersection of blue and red curves (left and right hand sides of the equation (\ref{sa1}) correspondingly). It is denoted by a red point which moves with the changing flux $\Phi(t)$.

Graph (ii) shows how the superconducting phase differences $\varphi^*,\;\delta^*$ across the studied and the reference JJ depend on the applied flux $\Phi$. Blue curve represents $\varphi^*(\Phi)$ and red curve represents $\delta^*(\varphi)$. Blue and red dots correspond to the $\Phi(t)$ value. 

Graph (iii) shows the critical current of the SQUID shifted by the critical current of the reference JJ and normalized by the critical current of the studied CPR (red curve) and a sinusoidal studied CPR (blue dashed curve). The red point corresponds to the $\Phi(t)$ value.

\subsection*{Supporting video 2}

Supporting video 2 consists of three animated graphs (i)-(iii) and describes an asymmetric SQUID that consists of two JJs with reference sinusoidal and studied skewed CPR
\begin{equation*}
    I_s^{JJ}(\delta) = \#\sin(\delta/2)\tanh\frac{\Delta\cos(\delta/2)}{2T}.
\end{equation*}
The critical current of the  studied JJ is 10 times smaller then of the reference JJ and $\delta/2T=10$.
In the graphs, the applied magnetic flux $\Phi(t)$ is parameterized with time.

Graph (i) shows the solution $\varphi^*(\Phi)$ of equation 
   \begin{equation}
        \partial_{\varphi}I_s^{REF}(\varphi) =-\partial_{\varphi} I_s^{JJ}(\varphi+2\pi\Phi/\Phi_0).
        \label{sa2}
    \end{equation}
The solution $\varphi^*$ is the intersection of red and blue curves (left and right hand sides of the equation (\ref{sa2}) correspondingly). It is denoted by a red point which moves with the changing flux $\Phi(t)$. Black dashed line shows the position of the critical phase of the reference JJ ($\varphi_c=\pi/2$).

Graph (ii) shows how the superconducting phase difference $\varphi^*$ across the reference JJ depend on the applied flux $\Phi$. The red dot correspond to the $\Phi(t)$ value. Black dashed line shows the position of the critical phase of the reference JJ ($\varphi_c=\pi/2$).

Graph (iii) shows the critical current of the SQUID shifted by the critical current of the reference JJ and normalized by the critical current of the studied CPR (red curve) and a skewed studied CPR (blue curve, equation (\ref{sa2})). The red point corresponds to the $\Phi(t)$ value.
\newpage
\section*{Supporting figures}
	\begin{figure*}[h!]
	\begin{center}
		\includegraphics[width=16cm]{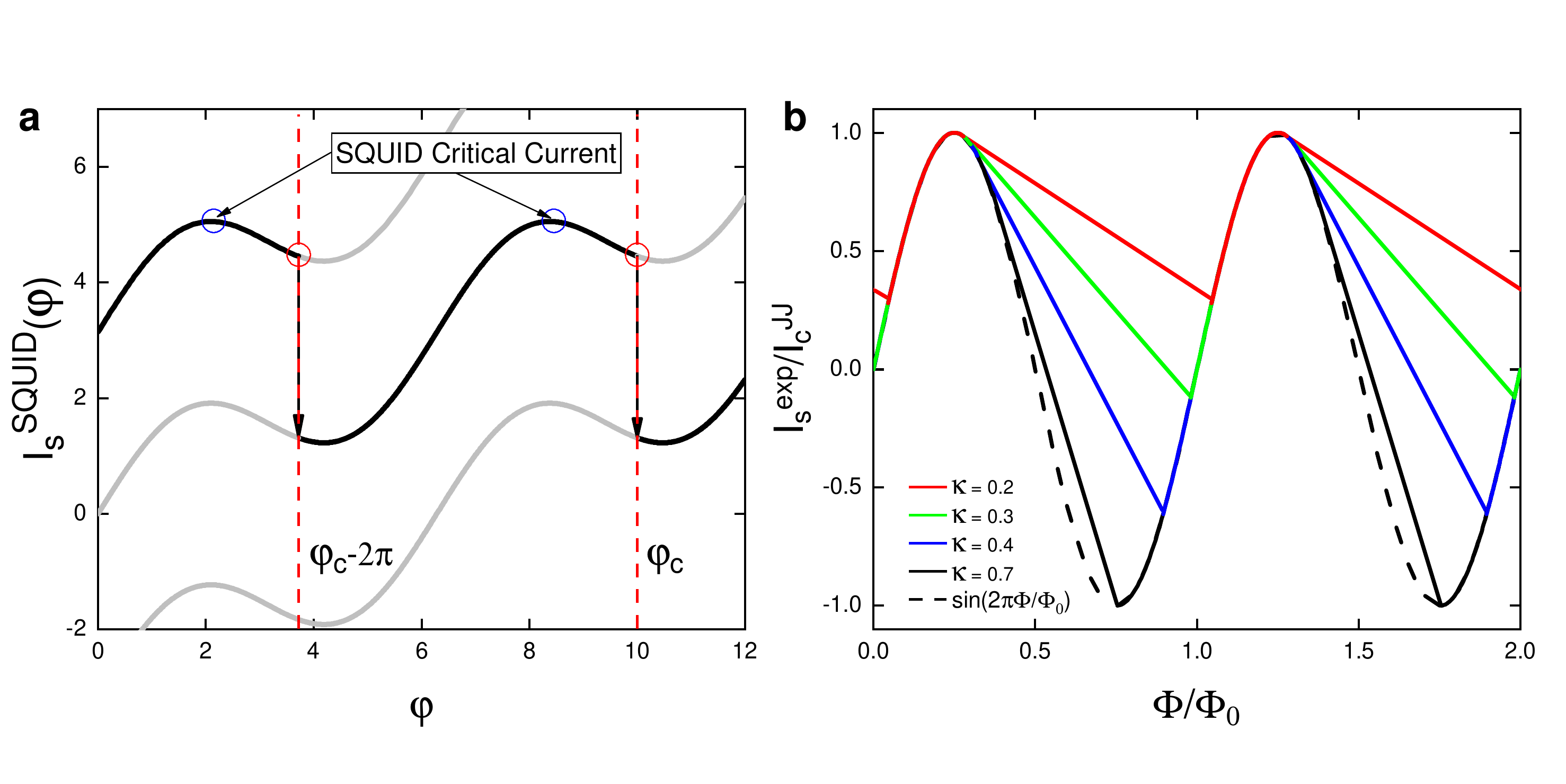} 
		\caption{\textbf{Model description.} (a) Supercurrent of the SQUID as a function of phase $\varphi$ over the nanobridge at the fixed magnetic flux for $\kappa=2/3$ (black line). Critical current is reached at the maximum value of supercurrent (blue circles) which is not at $\varphi_c$ (red circle). (b) ECPR, predicted by the model for different $\kappa$ parameters.
		}
		\label{m1}
	\end{center}
\end{figure*}
 \begin{figure*}[h!]
	\begin{center}
		\includegraphics[width=16cm]{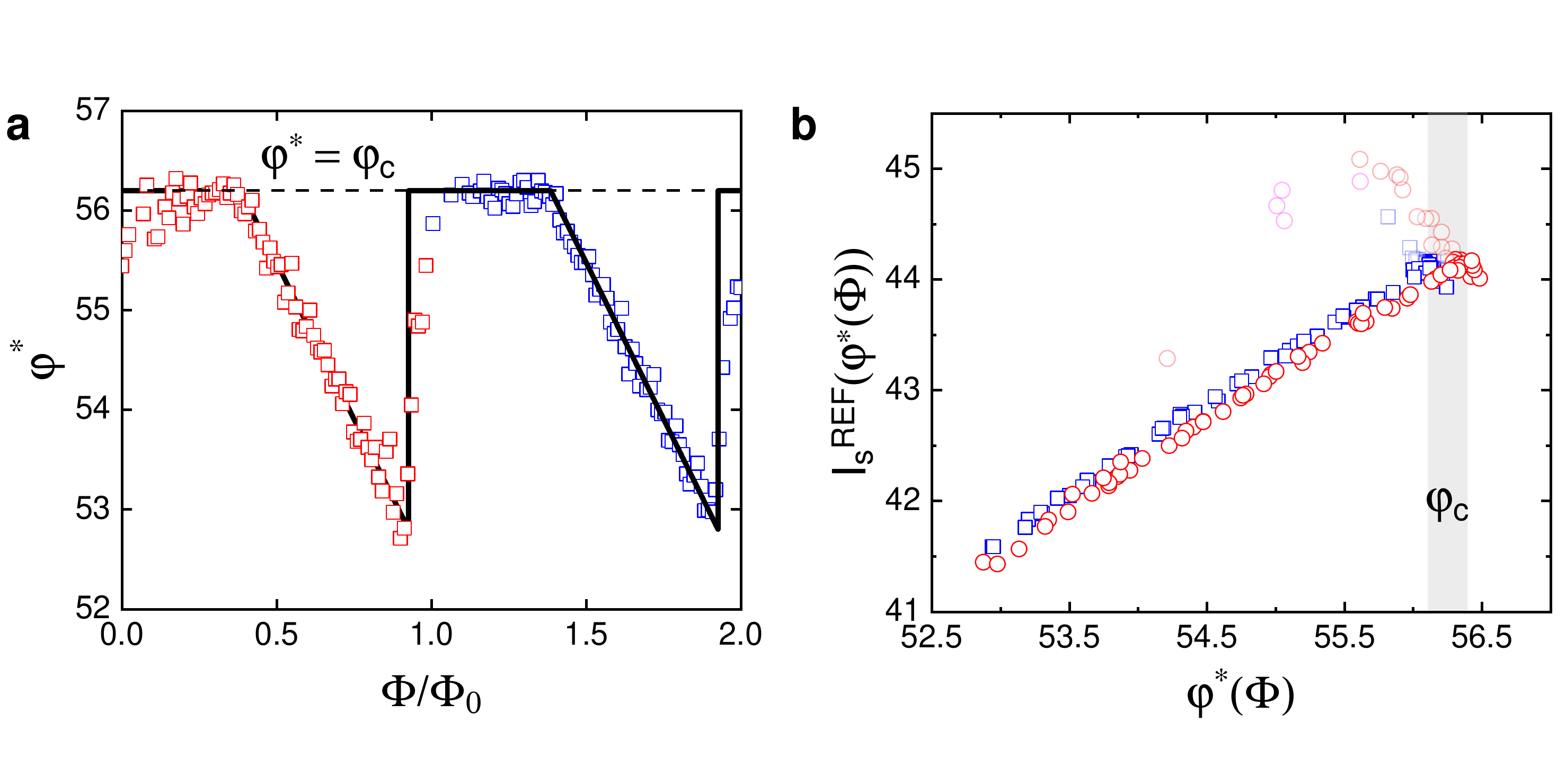} 
		\caption{\textbf{Phase and CPR  of the nanobridge.} (a) Extracted phase $\varphi^*(\Phi)$ of the nanobridge in the main device. The dashed line corresponds to the critical phase $\varphi_c$. The black line is the prediction of the model. (b) Nanobridges CPR extracted for the first two periods of oscillations (red and blue colors, respectively). 
		}
		\label{s6}
	\end{center}
\end{figure*}
 
\begin{figure}[h!]
	\begin{center}
		\includegraphics[width=13cm]{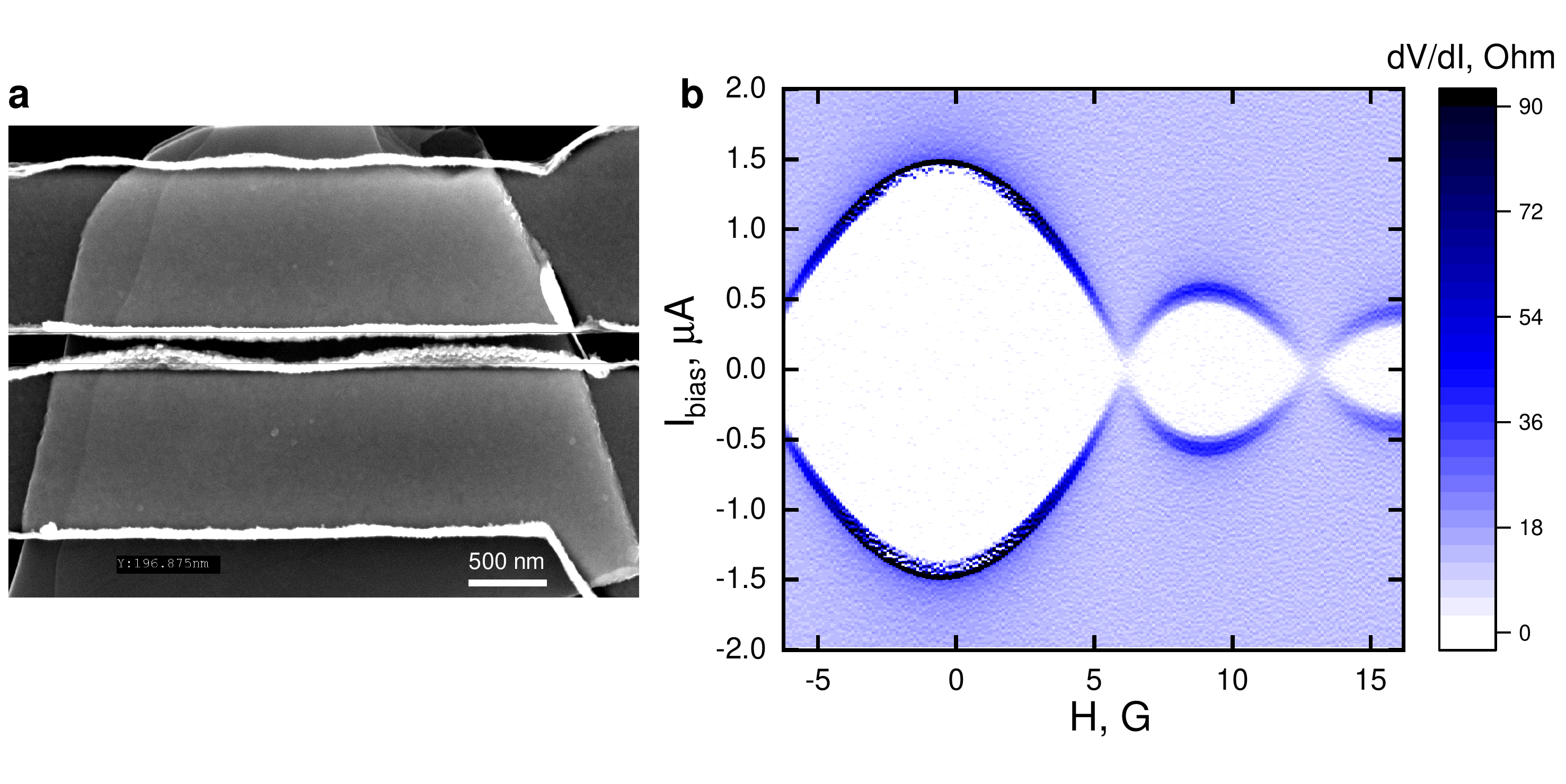} 
		\caption{\textbf{Single S-TI-S Josephson Junction} (a) SEM image of a Nb-Bi$_{2}$Te$_2$Se-Nb JJ. The distance between the leads is 200 nm. The thickness of the Nb leads is 100 nm. This JJ was fabricated using the same technological process, as described in the main text. (b) Color map of differential resistance as a function of the bias current and external magnetic field, obtained for the device, shown in (a). In Figure 1b of the main text, we rescaled the colormap from this Figure so that the positive critical current branch and the envelope of the SQUID oscillations were matched.
		}
		\label{s1}
	\end{center}
\end{figure}

\begin{figure}[h!]
	\begin{center}
		\includegraphics[width=15cm]{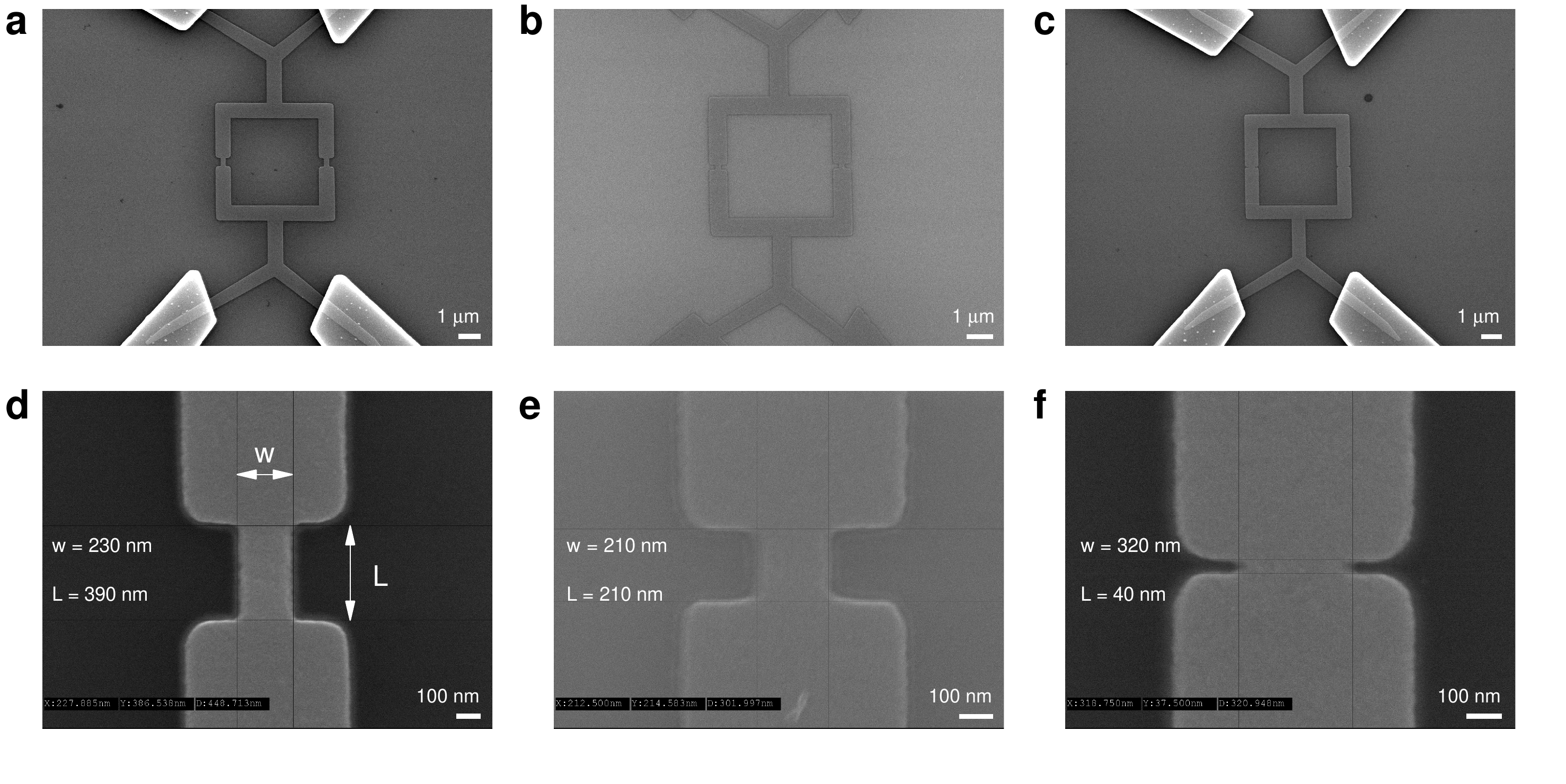} 
		\caption{\textbf{SEM images of symmetric SQUIDs.} (a)-(c) SQUID consisting of nanobridges with $L=390\;nm$, $L=210\;nm$ and $L=40\;nm$, respectively. These devices were fabricated using the same technological process, as described in the main text. The thickness of the Nb film is 20 nm (deposited by e-beam evaporation), and the thickness of the Nb contacts is 100 nm (deposited by magnetron sputtering). (d)-(f) Zoom of the left nanobridge from (a), (b) and (c), respectively. The width $w$ and the Lenght L of nanobridges are shown in the figures.
		}
		\label{nb1}
	\end{center}
\end{figure}

\begin{figure}[h!]
	\begin{center}
		\includegraphics[width=16cm]{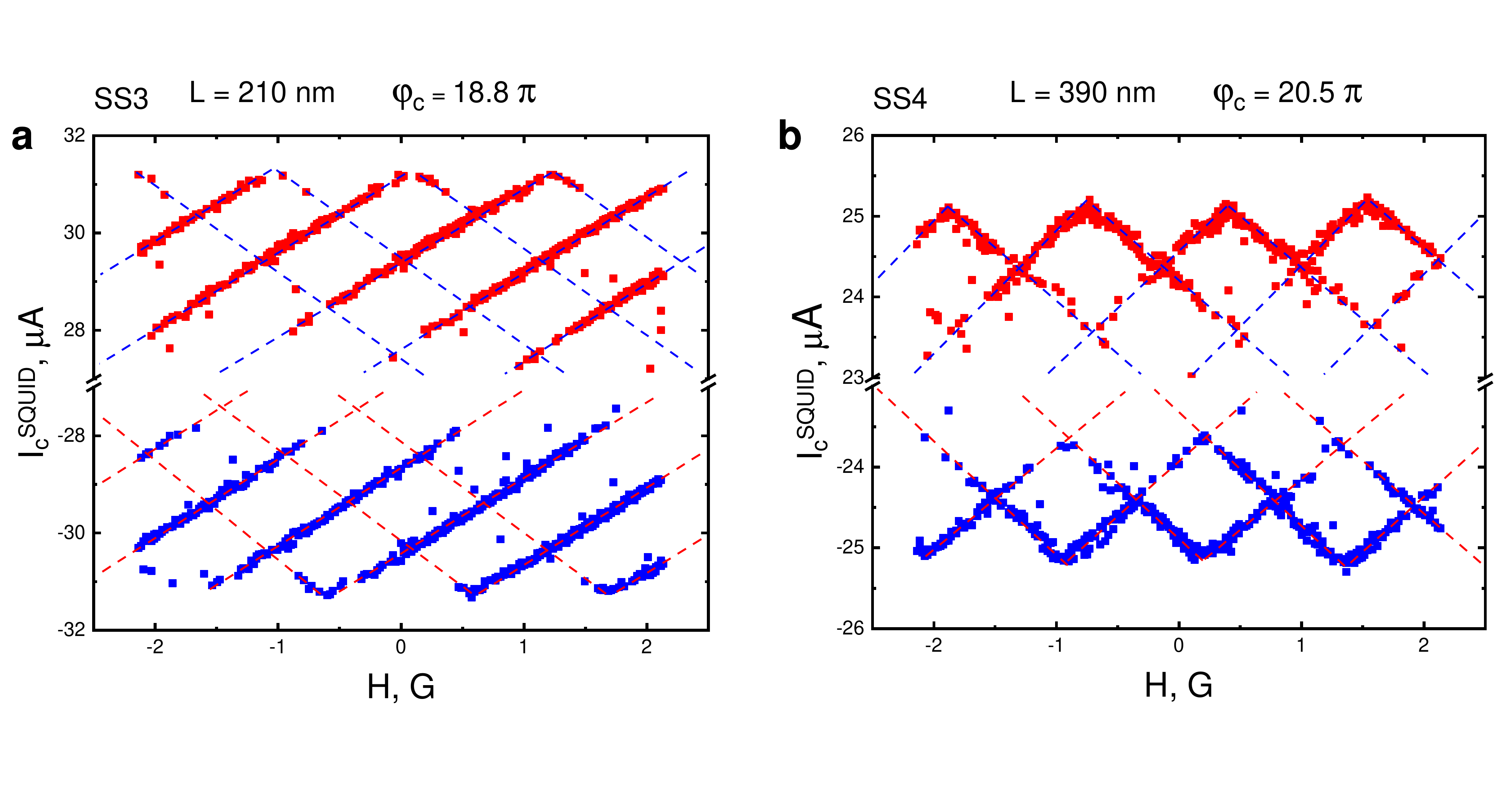} 
		\caption{\textbf{Symmetric SQUID oscillations of the devices from Substrate 1.} (a) and (b) show the $I_c(H)$ for nanobridges with $L=210\;nm$. Devices that correspond to (a) and (b) were fabricated in one technological process on the same substrate. We extracted critical phases of the nanobridges from the linear fit \cite{1,2}. As one can see, the values of the critical current and the critical phase are reproducible.
		}
		\label{nb2}
	\end{center}
\end{figure}

\begin{figure}[h!]
	\begin{center}
		\includegraphics[width=16cm]{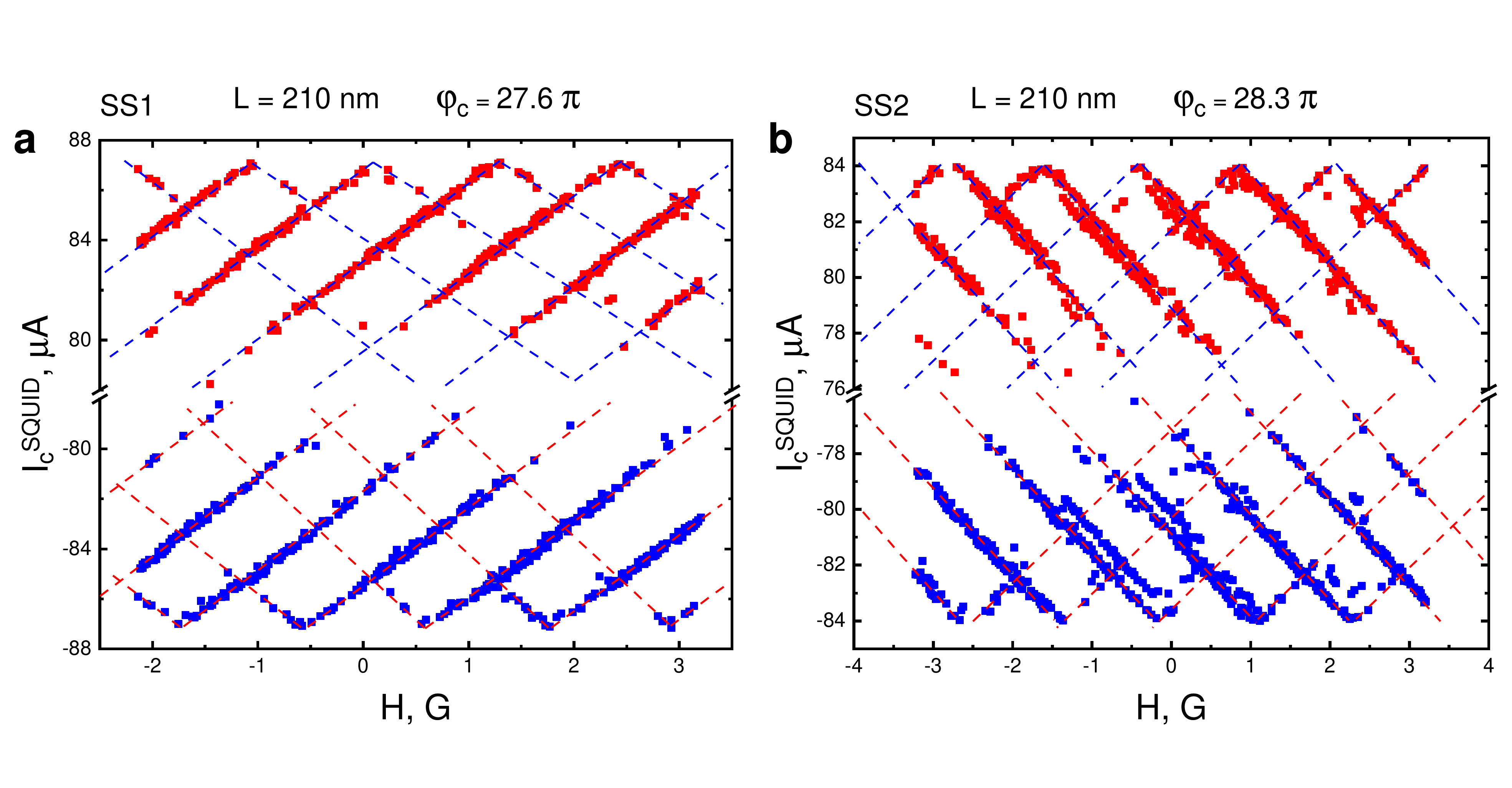} 
		\caption{\textbf{Symmetric SQUID oscillations of the devices from Substrate 2.} (a) The $I_c(H)$ for nanobridge with $L=210\;nm$. As one can see, the values of critical parameters are different, compared with devices from Figure S5. Therefore, the properties of the nanobridges are irreproducible between two fabrication cycles. (b) The $I_c(H)$ for nanobridge with $L=390\;nm$. As one can see, longer nanobridges have a smaller critical current and higher critical phase, as expected.
		}
		\label{nb3}
	\end{center}
\end{figure}

\begin{figure}[h!]
	\begin{center}
		\includegraphics[width=16cm]{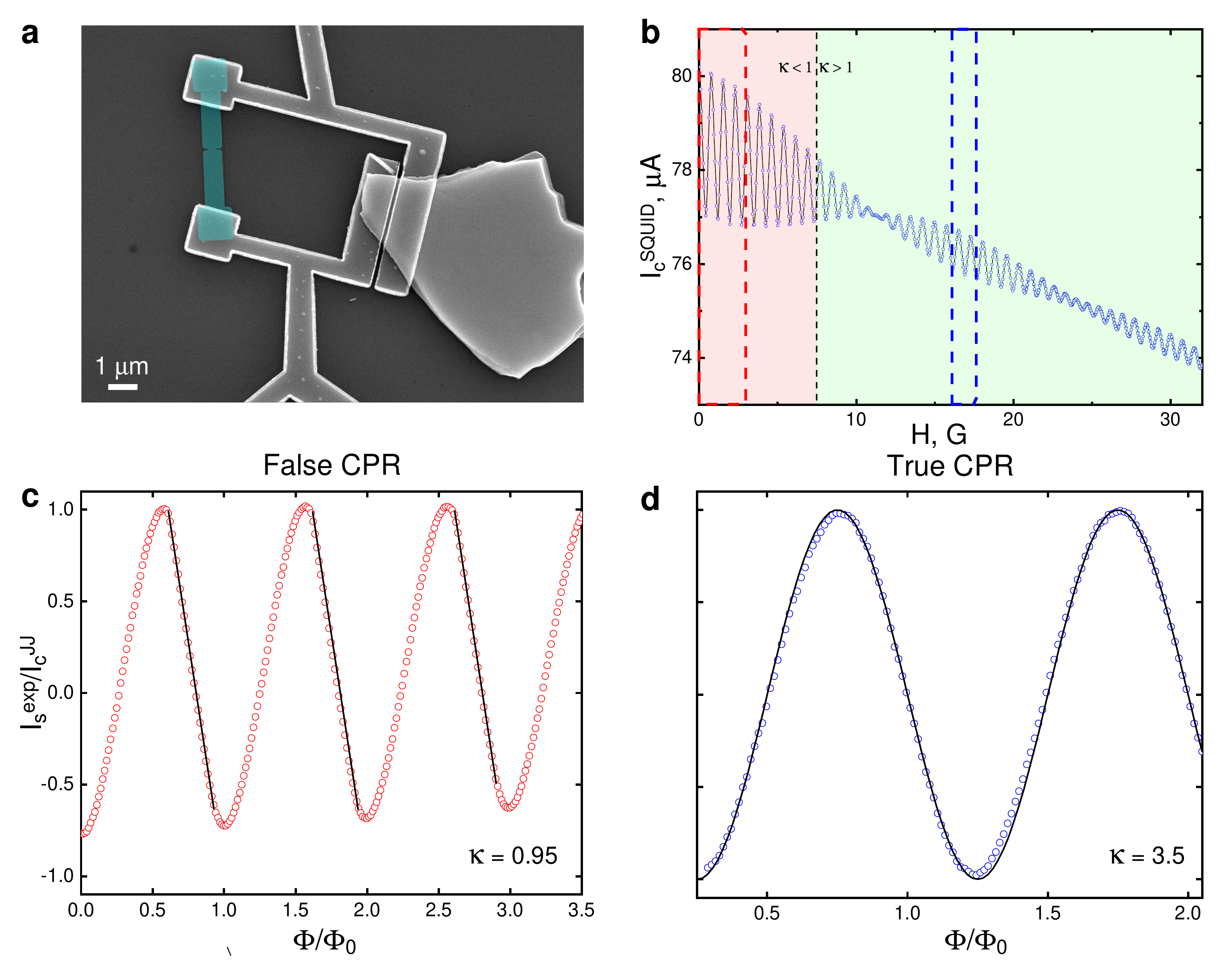} 
		\caption{\textbf{AS2 device.} (a) A false-colored SEM image of the AS2 device. The superconducting nanobridge (turquoise) is 35 nm long. The distance between leads in the S-TI-S JJ is 110 nm. (b) SQUID oscillations for AS2 device. Two measurement modes are observed. Red and green regions denote "False" and "True" CPR measurement regimes, respectively. The regions are separated by a black dashed line, which corresponds to $\kappa=I_c^{REF}/I_c^{JJ}\varphi_c=1$. (c) "False" CPR measurement regime from the red box, linear regions are indicated by black lines. (d) "True" CPR regime from the blue box. Almost sinusoidal CPR is observed.
		}
		\label{s2}
	\end{center}
\end{figure}

\begin{figure}[h!]
	\begin{center}
		\includegraphics[width=16cm]{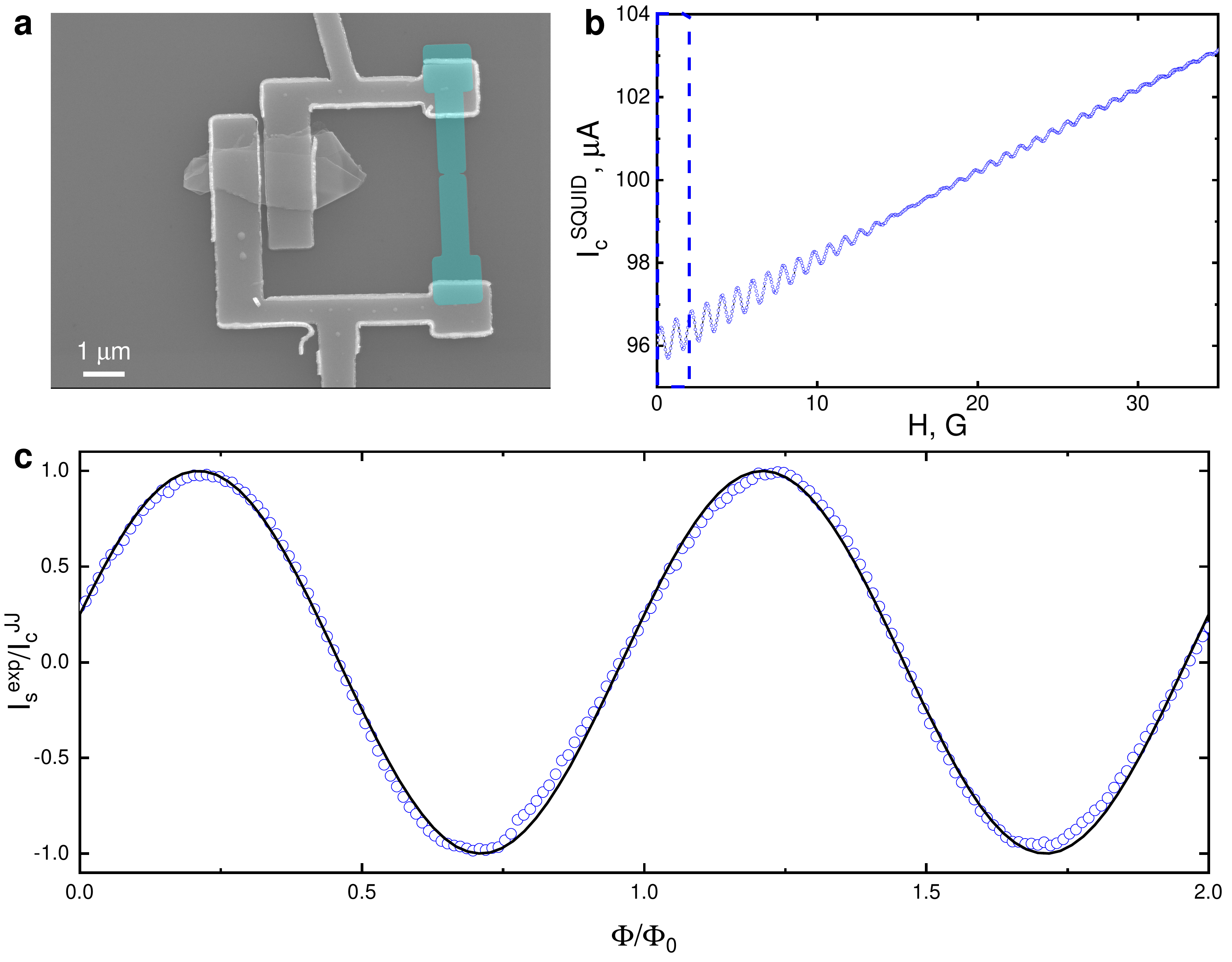} 
		\caption{\textbf{AS3 device.} (a) A false-colored SEM image of the AS3 device. The superconducting nanobridge (turquoise) is 35 nm long. The distance between leads of the S-TI-S JJ is 160 nm. (b) SQUID oscillations for device AS3, the device works entirely in the "True" measurement regime. (c) CPR from the blue box. The measured CPR is slightly skewed from the sinusoidal.
		}
		\label{s3}
	\end{center}
\end{figure}

\begin{figure}[h!]
	\begin{center}
		\includegraphics[width=16cm]{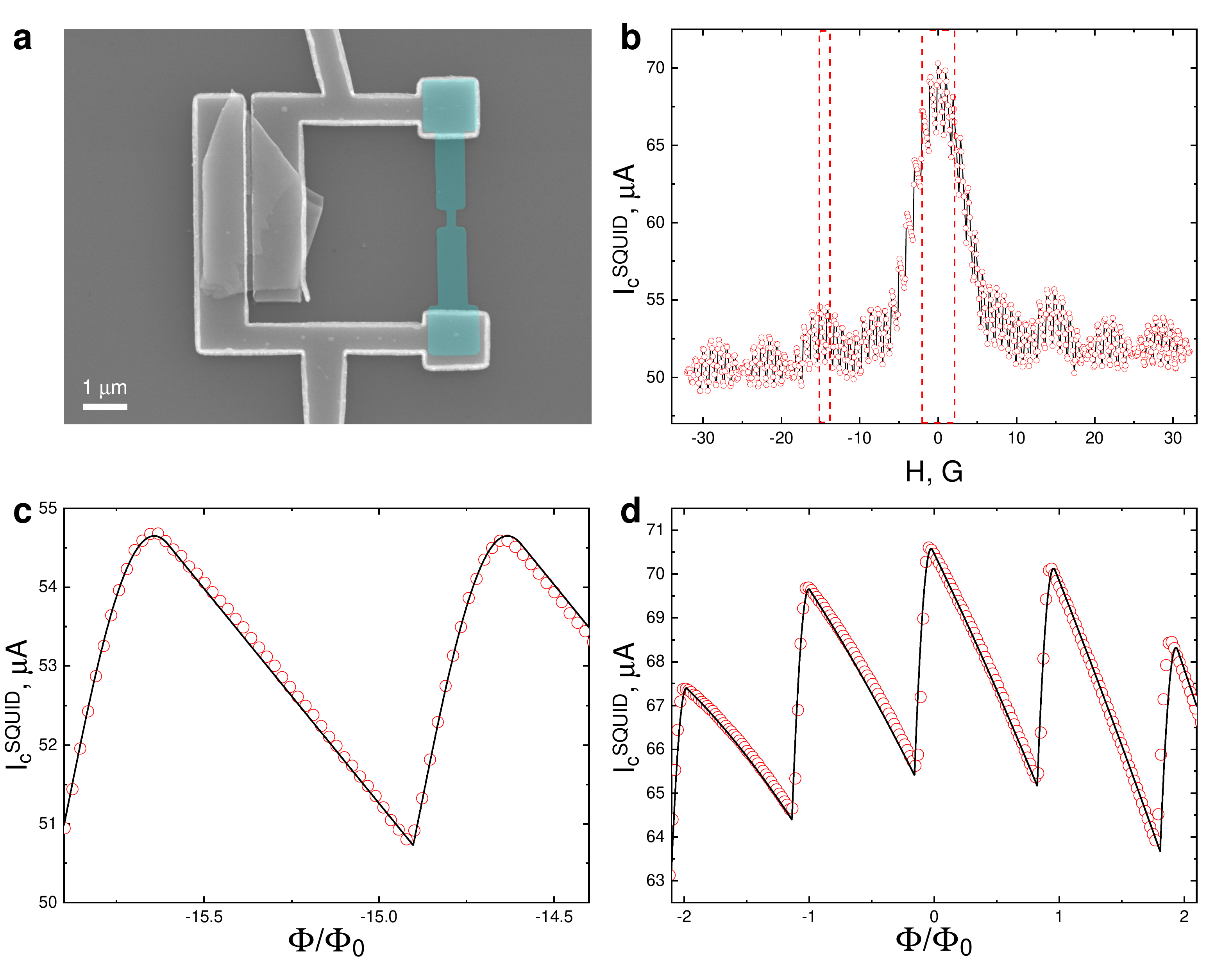} 
		\caption{\textbf{AS4 device.} (a) A false-colored SEM image of the AS4 device. The superconducting nanobridge (turquoise) is 390 nm long. The distance between leads of the S-TI-S JJ is 160 nm. (b) SQUID oscillations for the AS4 device, a pronounced peak in the zero-field region originates from the nanobridge's critical current. (c) SQUID oscillations in the high-field region (left red box in (b)). A black line is the fit with the model. (d) SQUID oscillations in small fields (right red box in (b)). A black line is the fit with the model.
		}
		\label{s4}
	\end{center}
\end{figure}